\documentclass[twocolumn,pra,aps,superscriptaddress,longbibliography]{revtex4-2}
\usepackage{amsmath}
\usepackage{amssymb}
\usepackage{amsfonts}
\usepackage[dvips]{graphicx}
\usepackage{subfigure}
\usepackage{dcolumn}
\usepackage{txfonts}
\usepackage{bm}
\usepackage{makeidx}
\usepackage{color}
\usepackage{mathtools}
\usepackage{threeparttable}
\usepackage[colorlinks,linkcolor=blue,anchorcolor=blue,citecolor=blue,urlcolor=blue]{hyperref}
\usepackage{lipsum}
\usepackage{braket}

\usepackage{mathrsfs}

\begin{document}

\title{Quantum Beam Splitter as a Quantum Coherence Controller}

\author{Li-Ping Yang}
\affiliation{Center for Quantum Sciences and School of Physics, Northeast Normal University, Changchun 130024, China}

\author{Yue Chang$^*$}
\affiliation{Beijing Automation Control Equipment Institute, Beijing, China} \affiliation{Quantum Technology R$\&$D Center of China Aerospace Science and Industry Corporation, Beijing, China}
\email{yuechang7@gmail.com}

\begin{abstract}
We propose a quantum beam splitter (QBS) with tunable reflection and transmission coefficients. More importantly, our device based on a Hermitian parity-time ($\mathcal{PT}$) symmetric system enables the generation and manipulation of asymmetric quantum coherence of the output photons. For the interference of two weak coherent-state inputs, our QBS can produce anti-bunched photons from one output port and bunched photons from the other, showcasing high parity asymmetry and strong coherence control capabilities. Beyond the Hong-Ou-Mandel effect, perfect photon blockade with vanishing $g^{(2)}(0)$ is achievable in two-photon interference. These striking effects of the QBS fundamentally arise from the parity-symmetry-breaking interaction and the quantum interference between the photon scattering channels. Our results could inspire novel applications and the development of innovative photonic devices for the manipulation of weak quantum light.
\end{abstract}

\maketitle
\section{Introduction}
A conventional beam splitter (CBS), which separates input light and transports its modes into two output ports, constitutes one of the key components in optical interferometers~\cite{hariharan2003optical,demkowicz2015quantum}. Following the development of its quantum theory~\cite{Fearn1987quantum,PRASAD1987quantum,Ou1987relation}, the CBS has found widespread applications such as linear optical quantum computing~\cite{knill2001scheme,Kok2007RMP,zhong2020quantum}, quantum imaging~\cite{Pittman1995,lemos2014quantum,Ndagano2022microscopy,moreau2019imaging}, and quantum sensing~\cite{dowling2008quantum,ligo2011gravitational,lloyd2008enhanced}. Usually, the CBS responds symmetrically to inputs from both sides, and its reflection and transmission coefficients determined by single-photon scattering processes remain fixed. Over the last decade, tunable polarization-independent beam splitters (BSs)~\cite{Ma2011tunable}, as well as tunable polarization and frequency BSs~\cite{Zhu2016Tunable,Wang2020Tunable,Hu2021onchip,Chang2021lowloss}, have been developed for photonic quantum information processing.  However, the CBS, being a linear optical device, lacks the capability to alter the statistical properties of the incident light. To generate photons with nontrivial quantum statistics, nonlinearity is required~\cite{Tian1992quantum,leo1994possibility,imamoglu1997strongly,birnbaum2005photon,faraon2008coherent,reinhard2012strongly,Snijders2018Observation,Bin2020NPhoton}. Based on media-induced photon-photon interaction, photon-number-conserving~\cite{Pezze06nonlinear} and non-conserving~\cite{fang2016experimental,PRA09controlled} nonlinear BSs have been proposed. The single-atom-based nonlinear BS has also been employed to manipulate the Hong-Ou-Mandel (HOM) interference~\cite{Oehri2015tubable,Roulet2016two}. Nevertheless, a BS with the capacity to control asymmetric photon transport and manipulate quantum coherence has not been fully explored. 


In this letter, we introduce a quantum beam splitter (QBS) based on a Hermitian $\mathcal{PT}$-symmetric system~\cite{Bender1998real,bender1999PT,ruter2010observation} that preserves the reciprocity at the single photon level~\cite{sakurai1995modern}. However, the broken parity symmetry results in non-reciprocity during multi-photon processes. The proposed QBS is designed to control photon transport and manipulate the quantum coherence of output photons. The reflection and transmission coefficients of the QBS can be finely tuned within a frequency range constrained by the spontaneous decay rates of the system. More importantly, anti-bunched and bunched photons can emerge in opposite output ports in a well-controlled manner. In the interference of two Fock-state photons, the $\mathcal{PT}$ symmetry of our QBS maintains equal output probabilities at the two output ports, yet the broken parity symmetry leads to distinct statistics as characterized by the second-order correlation function $g^{(2)}(\tau)$. Using state-of-the-art circuit QED technology~\cite{gu2017microwave,RMP21circuit,abdo2013nondegenerate,Zeytino2015microwave}, our QBS possesses the potential for seamless chip integration.

Over the past decade, substantial efforts have been dedicated to the development of lossless magnetic-field-free nonreciprocal devices~~\cite{yu2009complete,fan2012all,peng2014parity,sounas2017non,cao2017experimental,shi2015limitations,shen2016experimental}. Prior works have primarily focused on the non-reciprocal properties in the first-order coherence of photons, i.e., the non-reciprocal reflection and transmission coefficients~\cite{Scheucher16quantum,kim2015non,metelmann2015noneciprocal,Chen2022Nonreciprocal}. Recently, Doppler-effect-based non-reciprocal second-order coherence of output photons has started to receive attention~\cite{huang2018nonreciprocal,tang2022quantum,Shen2020Nonreciprocal,Jing2021nonreciprocal,Wang2019Nonreciprocal}, where photons input from opposite directions experience different transition frequencies of the scatters~\cite{wang2013optical,wu2014nonhermitian,maayani2018flying}. In contrast, both asymmetric single-photon transport and asymmetric multi-photon correlations in our QBS arise solely from the symmetry-breaking phase of atom-atom interaction and quantum interference between photon scattering channels~\cite{xu2015optical,xu2016nonreciprocal,bernier2017nonreciprocal}. This path-interference mechanism can also be extended to manipulate the higher-order coherence of photons. These findings have the potential to stimulate novel experiments and the creation of innovative nonreciprocal quantum devices.  

The article is organized as follows. In Sec.~\ref{sec2}, we present the theoretical model of our proposed QBS. In Sec.~\ref{sec3}, we give the analytical results of the reflection and transmission coefficients of the QBS. The controllable asymmetric second-order coherence of output photons is demonstrated in Sec.~\ref{sec4}. We reveal the two-photon interference beyond the traditional HOM effect in Sec.~\ref{sec5}. The implementation of our QBS and conclusions are summarized in Sec.~\ref{sec6}. The symmetry analysis of our system and the detailed multi-photon scattering approach used to evaluate the relevant quantities are provided in the Appendices.

\begin{figure}
\includegraphics[width=8cm]{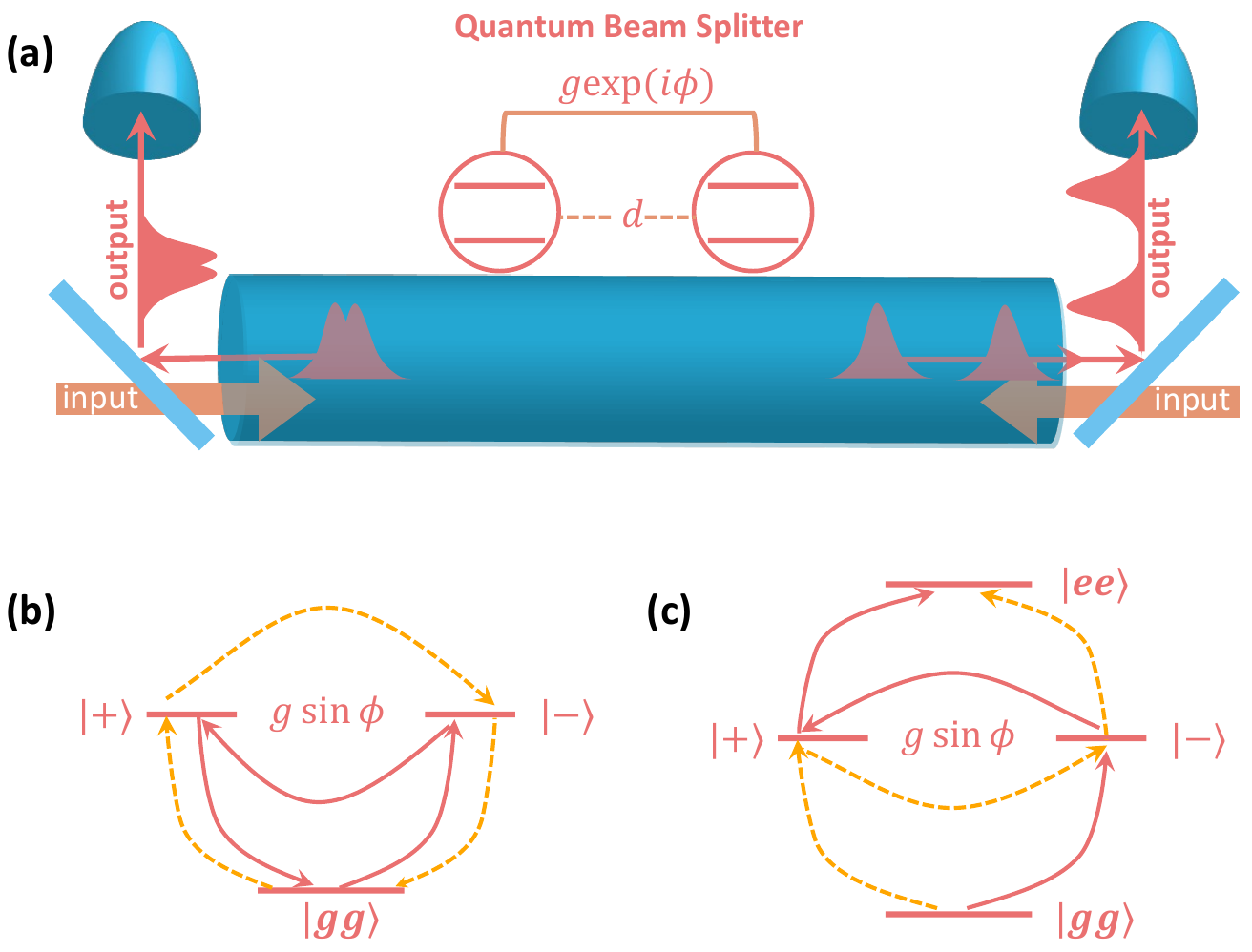}
\caption{\label{fig:1}
(a) The quantum beam splitter comprises a one-dimensional waveguide and two interacting atoms at a distance $d$. The two-level atoms have a ground state $|g\rangle$ and an excited state $|e\rangle$. When the atom-atom coupling phase $\phi$ is not integer multiples of $\pi$, both the parity and time-reversal symmetries of the system are broken. This leads to distinct statistical characteristics of the output photons from the left and right ports, representing a substantial departure from the behavior of the conventional beam splitters. (b) Quantum interference between two single-photon scattering channels. Here, $|\pm\rangle = (|eg\rangle \pm |ge\rangle )/\sqrt{2}$ are the positive and negative parity single-excitation states. The interference between the red-solid and dashed yellow single-photon scattering channels leads to a phase difference in the forward and backward transmission coefficients. (c) Quantum interference between two double-excitation channels. Similar interference also occurs between the de-excitation channels (not shown). The interference between the red-solid and dashed yellow two-photon scattering channels results in asymmetric correlation functions for the two output ports.}
\end{figure}

\section{Model\label{sec2}}
Our QBS consists of a bidirectional waveguide and two interacting two-level atoms as depicted in Fig.~\ref{fig:1} (a). In a rotating frame with respect to the atomic transition frequency $\omega_0$, the Hamiltonian of the whole system can be decomposed into three parts $\hat{H}=\hat{H}_a + \hat{H}_p +\hat{H}_{\rm int}$ (see Appendix \ref{Appendix1}). There exists a direct interaction between the two atoms 
\begin{equation}
\hat{H}_a = ge^{i\phi}\hat{\sigma}_{1}^{\dagger}\hat{\sigma}_{2}+ge^{-i\phi}\hat{\sigma}_{2}^{\dagger}\hat{\sigma}_{1}, \label{eq:atom_interaction}
\end{equation}
with atomic ladder operator $\hat{\sigma}=|g\rangle\langle e|$ and tunable coupling strength $g$ and phase $\phi$. The Hamiltonian of the wave-guide photons is given by $\hat{H}_p =\int k(\hat{b}^{\dagger}_{k,r}\hat{b}_{k,r}-\hat{b}^{\dagger}_{k,l}\hat{b}_{k,l})dk$ with bosonic operators $\hat{b}_{k,r}$ and $\hat{b}_{k,l}$ for right- and left-moving modes with momentum $k$~\cite{shi2011two,shen2005coherent,shen2005prl}. The atom-photon interaction $\hat{H}_{\rm int} = \sum_{i=1,2}\eta\int dk\hat{\sigma}_{i}\left[\hat{b}_{k,r}^{\dagger}e^{-i( k_{0}+ k)x_{i}}+\hat{b}_{k,l}^{\dagger}e^{i( k_{0}- k)x_{i}}\right]+{\rm H.c.}$ is of Jaynes–Cummings type, where $\eta$ is the interaction strength, the wavenumber $k_0$ corresponds to $\omega_0$, and the positions of the two atoms are $x_1=-d/2$ and $x_2=d/2$.  

The non-vanishing phase $\phi$ in interaction $\hat{H}_a$ breaks both the parity and time-reversal symmetries of this Hermitian system while preserving its $\mathcal{PT}$-symmetry (see Appendix~\ref{Appendix1}). This symmetry-breaking atom-atom interaction can be achieved in a circuit-QED system by modulating the qubit frequency and qubit-qubit coupling coefficient, as demonstrated in Appendix~\ref{Appendix6}. Phase-dependent atom-atom coupling has also been utilized to achieve non-reciprocal single-photon transport~\cite{wang2019phase,zhou2023chiral} and non-reciprocal single-photon blockade~\cite{lu2021plasmonic}. However, the $\mathcal{PT}$-symmetry of our BS maintains the reciprocity of single-photon reflectance and transmittance~\cite{sakurai1995modern}. 
The broken parity symmetry of the system, combined with we later revealed quantum interference, gives rise to asymmetric multi-photon scattering amplitudes and distinctive statistical properties of the output photons.

When coherent inputs (i.e., two monochromatic laser drivings) are applied to both sides, the reflectance, transmittance, and multi-time correlations of output photons can be obtained from the master equation together with the input-output formalism (see Appendix~\ref{Appendix2}). Under the Markov limit~\cite{Shi2015multiphoton}, the master equation for the density matrix $\rho_a$ of the atoms is given by
\begin{equation}
\partial_t\rho_a =-i[\hat{H}^{\prime}_a+\hat{H}_{\rm drive},\rho_a] +\sum_{ij}\Gamma_{ij}\left(2\hat{\sigma}_{i}\rho_{a}\hat{\sigma}_{j}^{\dagger}-\left\{\rho,\hat{\sigma}_{j}^{\dagger}\hat{\sigma}_{i}\right\}\right).
\end{equation}
The waveguide photons induce the spontaneous decay of atoms and modify atom-atom coupling $ge^{i\phi}$ to $ge^{i\phi}+\Gamma\sin\theta$ in Hamiltonian $\hat{H}^{\prime}_a$ (please refer to Appendix~\ref{Appendix2}), where $\Gamma = 2\pi |\eta|^2$, $\theta = k_0 d$, and $\Gamma_{ij}=\Gamma \cos [k_0(x_i-x_j)]$. The coherent drivings are described by  $\hat{H}_{\rm drive} = \sum_{i} \left[\Omega_r \hat{\sigma}_{i}^{\dagger}e^{i(k_0+p)x_i} + \Omega_l \hat{\sigma}_{i}^{\dagger}e^{-i(k_0-p)x_i}+ {\rm h.c.}\right]$ with pumping amplitudes $\Omega_{r}$ and $\Omega_{l}$ for   modes with wavenumber $k_0+p$ (right-moving) and $-(k_0 -p)$ (left-moving), respectively. In the weak-driving limit, the master equation can be related to few-photon scattering problems. Consequently, the characteristics of the output photons can be analytically assessed using the scattering method ~\cite{Shi2015multiphoton,Caneva2015Quantum,Chang2016deterministic}, with the scattering matrix determined by an effective non-Hermitian Hamiltonian $\hat{H}_{\rm eff}=\hat{H}^{\prime}_{a}-i\sum_{ij}\Gamma_{ij}\hat{\sigma}_{j}^{\dagger}\hat{\sigma}_{i}$.

To reveal the interference between the scattering channels, we introduce single-excitation states $|\pm\rangle  = \left(|eg\rangle \pm |ge\rangle\right)/\sqrt{2}$ with positive and negative parities, respectively. Utilizing ladder operators $\hat{\sigma}_{\pm}=(\hat{\sigma}_{1}\pm\hat{\sigma}_{2})/\sqrt{2}$, the effective non-Hermitian Hamiltonian is re-expressed as
\begin{equation}
\hat{H}_{{\rm eff}}=-i\left(\alpha_{+}\hat{\sigma}^{\dagger}_{+}\hat{\sigma}_{+}+\alpha_{-}\hat{\sigma}^{\dagger}_{-}\hat{\sigma}_{-}\right)-i\beta\left(\hat{\sigma}^{\dagger}_{+}\hat{\sigma}_{-}-\hat{\sigma}^{\dagger}_{-}\hat{\sigma}_{+}\right),
\end{equation}
with $\alpha_{\pm} =\Gamma(1\pm e^{i\theta})\pm ig\cos\phi$ and $\beta =g\sin\phi$. The dissipation terms are present only in the diagonal elements (see Appendix~\ref{Appendix3}), causing photon emission from states $|\pm\rangle$. The off-diagonal elements contain the parity-breaking term (odd power of $\sin\phi$), which leads to coherent transitions between states $|\pm\rangle$ and the interference of scattering channels, as depicted in Fig.~\ref{fig:1} (b) and (c). The phase $\phi$ can be used to control these quantum interference processes, thereby manipulating the transport and statistical properties of photons as shown in the following.

\section{Reflection and transmission coefficients\label{sec3}} 
The CBS is fully characterized by its reflection and transmission coefficients. These coefficients of our QBS are determined by single-photon scattering processes and can be measured via the output signal resulting from a weak continuous-wave coherent pump in experiments. For instance, the reflection and transmission coefficients for the right-moving input with momentum $k_0+p$ can be obtained through the relations $r_{r\rightarrow l} (p) =\langle \hat{b}_{{\rm out},l}(t)\rangle/\langle \hat{b}_{{\rm in},r}(t)\rangle$ and $t_{r\rightarrow r} (p) =\langle \hat{b}_{{\rm out},r}(t)\rangle/\langle \hat{b}_{{\rm in},r}(t)\rangle$ in the steady state. The input and output field operators are defined as $\hat{b}_{{\rm in},\lambda}(t)=\int dke^{-i\lambda k(t-t_i)}\hat{b}_{k,\lambda}(t_i)/\sqrt{2\pi}$ and $\hat{b}_{{\rm out},\lambda}(t)=\int dke^{-i\lambda k(t-t_i)}\hat{b}_{k,\lambda}(t_f)/\sqrt{2\pi}$~\cite{Caneva2015Quantum} (also see Appendix~\ref{Appendix2}), where $t_i \leq t+\lambda x_i\leq t_f$. We note that when $\lambda$ is not subscripted, 
$\lambda=\pm 1$ corresponding to right- and left-moving modes, respectively.

\begin{figure}
\includegraphics[width=8.5cm]{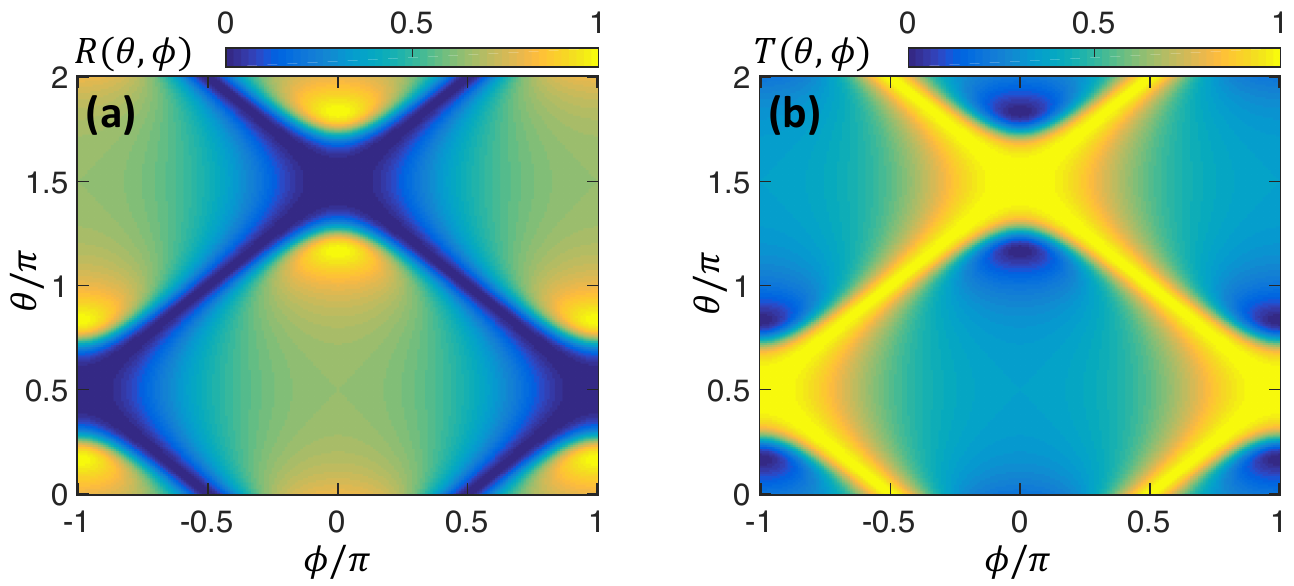}
\centering
\caption{\label{fig:S1} The reflectance $R(\theta,\phi)\equiv |r_{r\rightarrow l} (p)|^2$ (a) and transmittance $T(\theta,\phi)\equiv |t_{r\rightarrow r} (p)|^2$ (b) as functions of $\theta =k_0 d$ and $\phi$. Here, we consider the resonant input case with $p=0$. In our numerical simulation, we have taken the spontaneous decay rate as the unit of the frequency ($\Gamma=1$) and set the coupling strength as $g=\Gamma$.}
\end{figure}

Through symmetry analysis, some crucial properties of our BS can be deduced even without performing detailed calculations (see Appendix~\ref{Appendix1}). The Hamiltonian of our BS remains invariant under the $\hat{\mathcal{P}}\hat{\mathcal{T}}$ transformation. Consequently, the reflection coefficients for the input modes $\hat{b}_{-p,l}$ and $\hat{b}_{p,r}$ must be equal~\cite{sakurai1995modern}, i.e., $r_{l\rightarrow r} (-p) = r_{r\rightarrow l} (p)$. Even though the parity symmetry of our system is broken, the Hamiltonian under the space inversion transform simplifies as $\hat{\mathcal{P}}^{-1}\hat{H}(\phi)\hat{\mathcal{P}}=\hat{H}(-\phi)$. As a result, the transmission coefficients for the two modes $\hat{b}_{p,r}$ and $\hat{b}_{-p,l}$ are not equal; nevertheless, they are connected by the relation $t_{r\rightarrow r} (p,\phi) = t_{l\rightarrow l} (-p,-\phi)$. Similar relations
also apply to multi-photon scattering processes and higher-order correlation functions.

Using the recently developed multi-photon scattering method~\cite{Shi2015multiphoton,Caneva2015Quantum,Chang2016deterministic}, we can express the reflection and transmission coefficients in terms of the correlation functions of atomic operators (see Appendix~\ref{Appendix3})
\begin{align}
 r_{r\rightarrow l}(p) & = -\Gamma\sum_{ij}e^{ik_{0}(x_{i}+x_{j})}\left\langle gg\right|\hat{\sigma}_{i}\hat{M}\hat{\sigma}_{j}^{\dagger}\left|gg\right\rangle, \\
 t_{r\rightarrow r}(p) & = 1-\Gamma\sum_{ij}e^{-ik_{0}(x_{i}-x_{j})}\left\langle gg\right|\hat{\sigma}_{i}\hat{M}\hat{\sigma}_{j}^{\dagger}\left|gg\right\rangle, 
\end{align}
with $\hat{M}\!\!=\!\!(i\hat{H}_{\rm eff}\!-\!ip)^{-1}$. From the matrix elements of the effective Hamiltonian $\hat{H}_{\rm eff}$ in Eq.~(\ref{eq:Heff_mat}), we see that the states $|\pm\rangle$ determine the actual atomic decaying (photon emission) channels. We now re-express the scattering coefficients with operators $\hat{\sigma}_{\pm}$ and derive their analytical results
\begin{widetext}
\begin{align}
r_{r \rightarrow l}(p) & =-2\Gamma \left\langle gg\right|\left(\hat{\sigma}_{+}\cos\frac{\theta}{2}-i\hat{\sigma}_{-}\sin\frac{\theta}{2}\right)\hat{M}\left(\hat{\sigma}_{+}^{\dagger}\cos\frac{\theta}{2}-i\hat{\sigma}_{-}^{\dagger}\sin\frac{\theta}{2}\right)\left|gg\right\rangle,\label{eq:r1}\\
& =-\Gamma\left[2\left(M_{22}\cos^{2}\frac{\theta}{2}\!-\!M_{33}\sin^{2}\frac{\theta}{2}\right)\!-\!i(M_{23}\!+\!M_{32})\sin\theta\right]= 2i\Gamma\frac{\Gamma\sin\theta+p\cos\theta+g\cos\phi}{D(p)},\label{eq:rl1}\\
 t_{r\rightarrow r}(p)
& = 1-2\Gamma \left\langle gg\right|\left(\hat{\sigma}_{+}\cos\frac{\theta}{2}+i\hat{\sigma}_{-}\sin\frac{\theta}{2}\right)\hat{M}\left(\hat{\sigma}_{+}^{\dagger}\cos\frac{\theta}{2}-i\hat{\sigma}_{-}^{\dagger}\sin\frac{\theta}{2}\right)\left|gg\right\rangle,\label{eq:t1}\\
& = 1 -\Gamma\left[2\left(M_{22}\cos^{2}\frac{\theta}{2}+M_{33}\sin^{2}\frac{\theta}{2}\right)-i(M_{23}-M_{32})\sin\theta\right] = \frac{g^{2}-p^{2}+2g\Gamma e^{-i\phi}\sin\theta}{D(p)} 
,\end{align}\end{widetext}
where $D(p)=(\alpha_{+}-ip)(\alpha_{-}-ip)+\beta^{2}$ and the elements $M_{ij}$ are given in Eq.~(\ref{eq:M_mat}).
The other two scattering coefficients can be obtained similarly,
\begin{align}
r_{l \rightarrow r}(-p)=r_{r \rightarrow l}(p),\     t_{l\rightarrow l}(-p) = \frac{g^{2}-p^{2}+2g\Gamma e^{i\phi}\sin\theta}{D(p)}. 
\end{align}
The reflection and transmission coefficients satisfy two basic identities
\begin{equation}
\left|r_{r\rightarrow l}(p)\right|^2 + \left|t_{r\rightarrow r}(p)\right|^2  = \left|r_{l\rightarrow r}(-p)\right|^2 + \left|t_{l\rightarrow l}(-p)\right|^2 =1, \label{eq:rt_relation1}   
\end{equation}
and 
\begin{equation}
r^{*}_{r\rightarrow l}(p) t_{l\rightarrow l}(-p) + r_{l\rightarrow r}(-p) t^{*}_{r\rightarrow r}(p) =0, \label{eq:rt_relation2}   
\end{equation}
for lossless BSs due to the energy conservation law~\cite{Zeilinger1981general,loudon2000quantum}.

\begin{figure*}
\includegraphics[width=16cm]{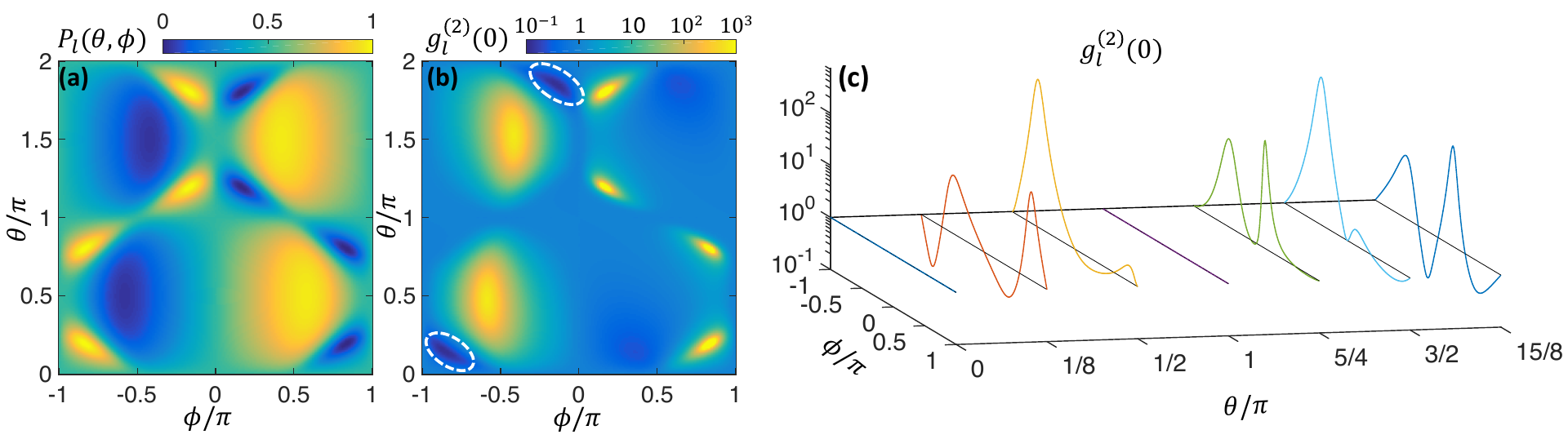}
\centering
\caption{\label{fig:2} The output probability (a) and  $g^{(2)} (0)$ function (b) of the photons from the left port for the interference of two resonant coherent-state inputs as functions of atomic distance $\theta = k_0 d$ and phase $\phi$. (c) Behavior of the $g^{(2)} (0)$ function for specific  $\theta$ values. The output probability and statistical characteristics for the right port can be obtained by substituting $\phi\rightarrow -\phi$ based on those of the left port. }
\end{figure*}

The analytical results above help us visualize the quantum interference in single-photon scattering processes. Equations (\ref{eq:r1}) and (\ref{eq:t1}) show that there exist four single-photon scattering channels: (1) $\left|gg\right\rangle \rightarrow\left|+\right\rangle \rightarrow|gg\rangle$; (2) $|gg\rangle \rightarrow\left|-\right\rangle \rightarrow |gg\rangle$; (3) $|gg\rangle \rightarrow\left|+\right\rangle \rightarrow\left|-\right\rangle \rightarrow |gg\rangle$; (4) $|gg\rangle \rightarrow\left|-\right\rangle \rightarrow\left|+\right\rangle \rightarrow |gg\rangle$. Due to $M_{23}=-M_{32}$, a perfect destructive interference occurs between path (3) and path (4) as shown in Fig.~\ref{fig:1} (b). This results in the complete cancellation of the parity-symmetry-breaking terms in the reflection coefficients. Consequently, the single-photon reflection processes are reciprocal  [see Fig.~\ref{fig:S1} (a)], as ensured by the $\mathcal{PT}$-symmetry. In contrast, a perfect constructive interference between these two paths arises in the transmission processes resulting in asymmetric coefficients $t_{r\rightarrow r}(p)\neq t_{l\rightarrow l}(-p)$. Nevertheless, the reciprocity of the transmittance is preserved in adherence to the probability conservation law, i.e., $\left|t_{r\rightarrow r}(p)\right|^2 = \left|t_{l\rightarrow l}(-p)\right|^2$ [see Fig.~\ref{fig:S1} (b)].

We now show that the symmetry-breaking phase $\phi$ can be used to manipulate the output probabilities of our QBS. Although the two reflection coefficients are identical, the two complex transmission coefficients have the same magnitude but different phases. Therefore, when considering the interference of the QBS with two identical weak coherent-state inputs, the output probability at its two ends $P_{\lambda} = \langle \hat{b}^{\dagger}_{{\rm out},\lambda}(t)\hat{b}_{{\rm out},\lambda}(t)\rangle/[\langle \hat{b}^{\dagger}_{{\rm in},l}(t)\hat{b}_{{\rm in},l}(t)\rangle+\langle \hat{b}^{\dagger}_{{\rm in},r}(t)\hat{b}_{{\rm in},r}(t)\rangle]$ could be significantly different, yet they are connected by the relationship $P_l (p,\phi)=P_r (p,-\phi)$. These two output probabilities can be expressed analytically as $P_l(p)=\left|r_{r\rightarrow l}(p)+t_{l\rightarrow l}(-p)\right|^2$ and  $P_r(p)=\left|r_{l\rightarrow r}(-p)+t_{r\rightarrow r}(p)\right|^2$, and they can be continuously tuned in practical applications. The probability $P_l$ as a function of atomic distance $\theta$ and the phase $\phi$ for resonant inputs with $p=0$ is presented in Fig.~\ref{fig:2} (a).

\section{Asymmetric $g^{(2)}$-functions\label{sec4}}
Different from the linear CBS, the coupling between light and atoms in our QBS mediates interactions among the incident photons, consequently altering their statistical characteristics depicted by the second-order coherence in the steady state
\begin{equation}
 g^{(2)}_{\lambda}(\tau)=\frac{\langle \hat{b}^{\dagger}_{{\rm out},\lambda}(t)\hat{b}^{\dagger}_{{\rm out},\lambda}(t+\tau)\hat{b}_{{\rm out},\lambda}(t+\tau)\hat{b}_{{\rm out},\lambda}(t)\rangle}{\langle\hat{b}^{\dagger}_{{\rm out},\lambda}(t)\hat{b}_{{\rm out},\lambda}(t)\rangle\langle\hat{b}^{\dagger}_{{\rm out},\lambda}(t+\tau)\hat{b}_{{\rm out},\lambda}(t+\tau)\rangle}.
\end{equation}
For the interference of two identical weak coherent-state inputs, the equal-time correlation of the photons exiting the left port of our QBS can be characterized by (see Appendix~\ref{Appendix4})
\begin{align}
&g_{l}^{(2)}(0) = \nonumber \\
&\frac{\left|ip\left[1\!-\!2r_{r \rightarrow l}(p)\!-\!2t_{l\rightarrow l}(-p)\!\right]\!+\!\Gamma[1\!+\!t_{l\rightarrow l}(-p)\!-\!t_{r\rightarrow r}(p)]\right|^{2}}{\left|r_{ r\rightarrow l}(p)+t_{l\rightarrow l}(-p)\right|^{4}|\Gamma-ip|^{2}}.\label{eq:g20l-1}
\end{align}
The $g^{(2)}$-function of the output photons from the right port differs from Eq.~(\ref{eq:g20l-1}), yet it can be determined using the relation $g_r^{(2)}(0,\phi)=g_l^{(2)}(0,-\phi)$. The interference between the two double-excitation paths [depicted in Fig.~\ref{fig:1} (c)] plays a key role in the asymmetric two-photon scattering processes, leading to non-reciprocal $g^{(2)}(0)$ functions. 

For the resonant input with $p=0$, the correlation function $g^{(2)}(0)$ ranges from $10^{-1}$ to $10^{3}$ as shown in Fig.~\ref{fig:2} (b), corresponding to various output probabilities.  Notably interesting regions are demarcated by white-dashed circles with relatively large output probabilities and concurrently small $g^{(2)}(0)\ll 1$. When $\theta = n\pi$ with integer $n$, the $g^{(2)}(0)$ function is the constant $1$ for photons coming out from both the left and right ports, as depicted in Fig.~\ref{fig:2} (c). When $\theta \neq n\pi$, the $g^{(2)}(0)$-function can be finely tuned via the phase $\phi$. Super-Poissonian and sub-Poissonian photons can be obtained simultaneously from the two respective output ports around $\phi =0$ for the case with the atomic distance corresponding to $\theta = \pi/8$.

\begin{figure}
\includegraphics[width=8.5cm]{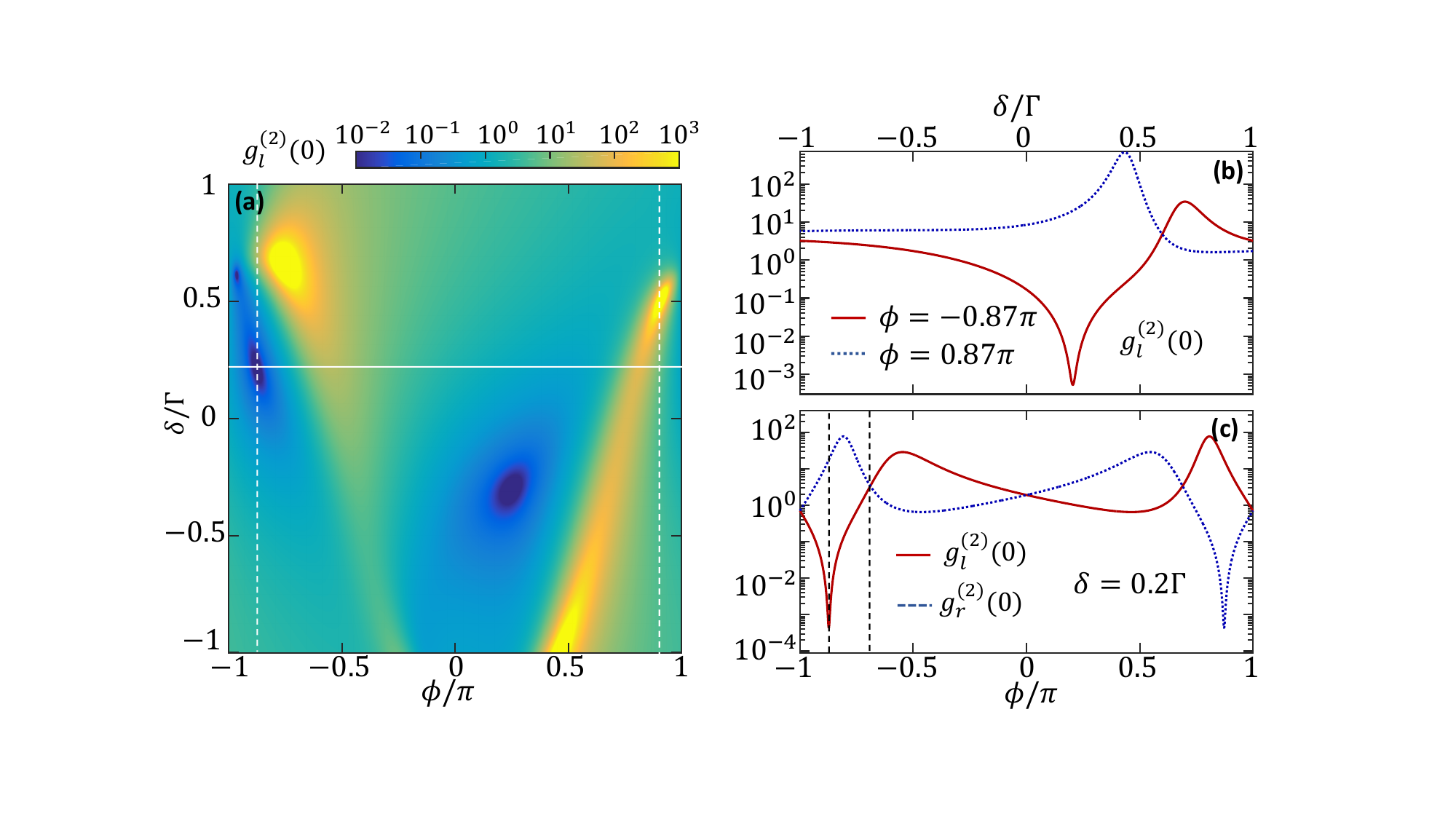}
\centering
\caption{\label{fig:3} (a) The influence of the detuning $\delta$ and phase $\phi$ on the correlation function $g^{(2)}_l(0)$. (b) The $g^{(2)}_r(0)$ as a function of detuning $\delta$ for $\phi = \pm 0.87\pi$. These two curves correspond to the two vertical white-dashed lines in panel (a). (c) The correlation functions for the left and right output ports with $\delta = 0.2\Gamma$ corresponding to the horizontal white-solid line in panel (a). We will examine the behavior  of $g^{(2)}(\tau)$ for the parameters $\phi=-0.68\pi$ and $\phi=-0.87\pi$ indicated by the two vertical dashed lines in Fig.~\ref{fig:4}. In this figure, the atomic distance is set as $\theta = \pi/8$. }
\vspace{-0.5cm}
\end{figure}

A wider range of $g^{(2)}(0)$ can be obtained by adjusting the atom-photon detuning $\delta$ (equivalently, the momentum $p$) as shown in Fig.~\ref{fig:3}. In panel (a), the correlation function $g^{(2)}_l(0)$ of photons exiting the left port is depicted against $\phi$ and $\delta$ with fixed $\theta=\pi/8$. For a more detailed illustration, we plot $g^{(2)}_l(0)$ as a function of $\delta$  in panel (b) for $\phi=\pm 0.87 \pi$ corresponding to the two vertical white-dashed lines in panel (a). To highlight the distinction in the photon statistics between the outputs from the left and right ports, we plot $g^{(2)}_l(0)$ and $g^{(2)}_r(0)$ with fixed $\delta =0.2\Gamma$ in panel (c). This confirms the identity $g^{(2)}_l (0,\phi) =g^{(2)}_r (0,-\phi) $ deduced from symmetry analysis. At $\phi = -0.87\pi$, the photons from the left port exhibit strong sub-Poissonian statistics with $g^{(2)}_l(0)\approx 4.3\times 10^{-4}$ and output probability $P_l\approx 83\%$, while those from the right port display strong super-Poissonian behavior with $g^{(2)}_r(0)\approx 17.9$ and output probability $P_r\approx 17\%$ [see Fig.~\ref{fig:S2} (a)]. This signifies a difference of more than 4 orders of magnitude in $g^{(2)}(0)$ for opposite directions. Unlike the previous approach utilizing different detuning for left- and right-moving photons~\cite{huang2018nonreciprocal}, this giant asymmetry in our QBS is solely attributable to quantum interference effects.

\begin{figure}
\includegraphics[width=8.5cm]{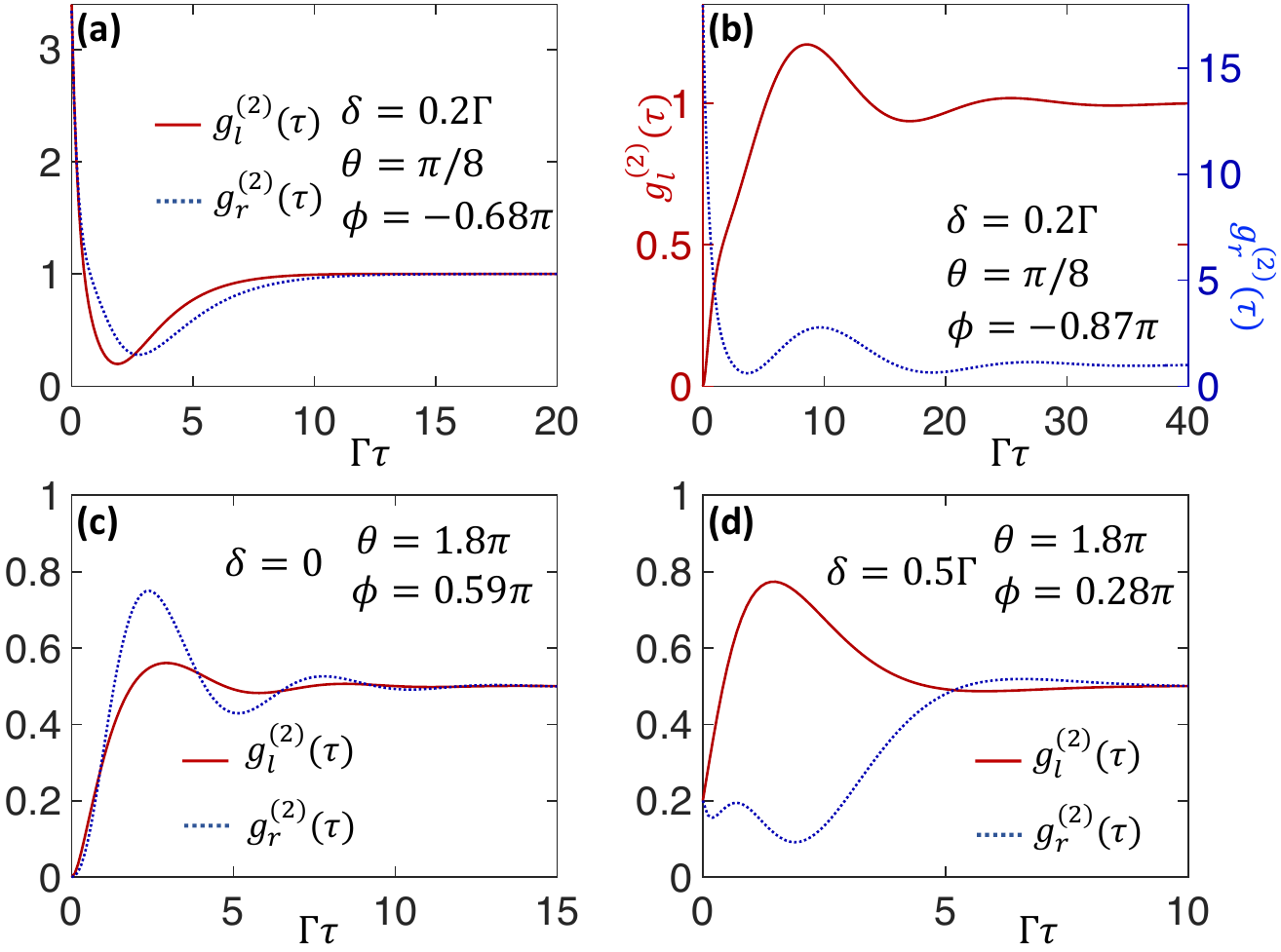}
\centering
\caption{\label{fig:4} Second-order correlation functions $g^{(2)} (\tau)$ of the output photons from the left and right ports. (a) and (b) depict the $g^{(2)}(\tau)$ for the interference of two coherent-state inputs. A double $y$-axis plot has been employed in (b). The left and right $y$-axes correspond to the functions $g^{(2)}_l$ and $g^{(2)}_r$, respectively. (c) and (d) showcase the $g^{(2)}(\tau)$ for the Hong-Ou-Mandel interference involving resonant and off-resonant input photons.}
\vspace{-0.5cm}
\end{figure}

To further examine the quantum coherence of the output photons, we investigate the $g^{(2)}(\tau)$ functions~\cite{zou1990photon,loudon2000quantum,Miranowicz2010testing}. The correlation functions of photons from the left and right output ports are also linked by the relation $g^{(2)}_l(\tau,\phi)=g^{(2)}_r(\tau,-\phi)$. Consequently, $g^{(2)}_l(0)$ and $g^{(2)}_r(0)$ must intersect at $\phi=0$ and $\phi=\pm\pi$ as shown in Fig.~\ref{fig:3} (c) and their corresponding $g^{(2)}(\tau)$ functions also overlap completely (not shown). At other points, the parity symmetry of the $g^{(2)}(\tau)$ functions is broken, i.e., $g_l^{(2)}(\tau)\neq g_r^{(2)}(\tau)$. For the accidental intersection point of $g^{(2)}(0)$ [such as $\phi = -0.68\pi$ in Fig.~\ref{fig:3} (c)], the output photons from both sides are bunched with $g^{(2)}(\tau)<g^{(2)}(0)$ as shown in Fig.~\ref{fig:4} (a). However, $g^{(2)}_l(\tau)$ differs from $g^{(2)}_r(\tau)$ slightly due to the interference between the de-excitation channels $|ee\rangle\rightarrow |gg\rangle$ (see Appendix\ref{Appendix4}), which are the inverse processes of Fig.~\ref{fig:1} (c). In Fig.~\ref{fig:4} (b), a more intriguing phenomenon is observed at $\phi = -0.87\pi$. The left outgoing sub-Poissionian photons are anti-bunched [$g^{(2)}_l(\tau)>g^{(2)}_l(0)$], while the right outgoing super-Poissionian photons are bunched [$g^{(2)}_r(\tau)<g^{(2)}_r(0)$].

\section{HOM interference\label{sec5}}
The two-photon HOM interference has important applications in quantum sensing and quantum state engineering. When two identical photons encounter a balanced $50:50$ CBS, they unfailingly emerge together at the same output port~\cite{HOM1987}. The HOM effect also manifests in our QBS when $P_r = P_l=1/2$. Unlike the scattering of a weak coherent-state input from one side~\cite{Shi2015multiphoton,Caneva2015Quantum,Chang2016deterministic}, the two-photon interference involving Fock-state inputs from two ports differs from coherent driving significantly. The master-equation approach cannot be applied in this context. We address this case using scattering theory and show that our QBS unveils intriguing effects beyond the conventional HOM effect.

For HOM interference of two input photons in the Fock-state $\hat{b}^{\dagger}_{r}(p)\hat{b}^{\dagger}_{l}(-p)|0\rangle$, the $g^{(2)}(0)$ functions of the output photons from the two ports are always the same (see Appendix~\ref{Appendix5})
\begin{equation}
g^{(2)}_l (0) =g^{(2)}_r (0) = \frac{1}{2}\left|1-\frac{\Gamma}{(\Gamma-ip)}\right|^2.
\end{equation}
This marks a significant difference from the interference of two coherent-state photons. Specifically, for the resonant inputs with $\delta = p =0$, the $g^{2}(0)$-function vanishes as depicted in Fig.~\ref{fig:4} (c), showing perfect photon blockade in the output photons~\cite{Oehri2015tubable}. In the non-resonant case shown in panel (d), the photons from opposite ports have the same $g^{(2)}(0)$ value, but their $g^{(2)}(\tau)$ functions behave differently. The photons outgoing from the left port display anti-bunching, whereas those from the right port exhibit a bunching effect within a small time scale $\tau<\Gamma^{-1}$. The two photons coexisting at one of the output ports of a CBS due to the HOM interference will exhibit a constant $g^{(2)}(\tau)=1/2$. This corresponds to the large-$\tau$ limit in our QBS, where the two photons are scattered independently.

\begin{figure}
\includegraphics[width=8.5cm]{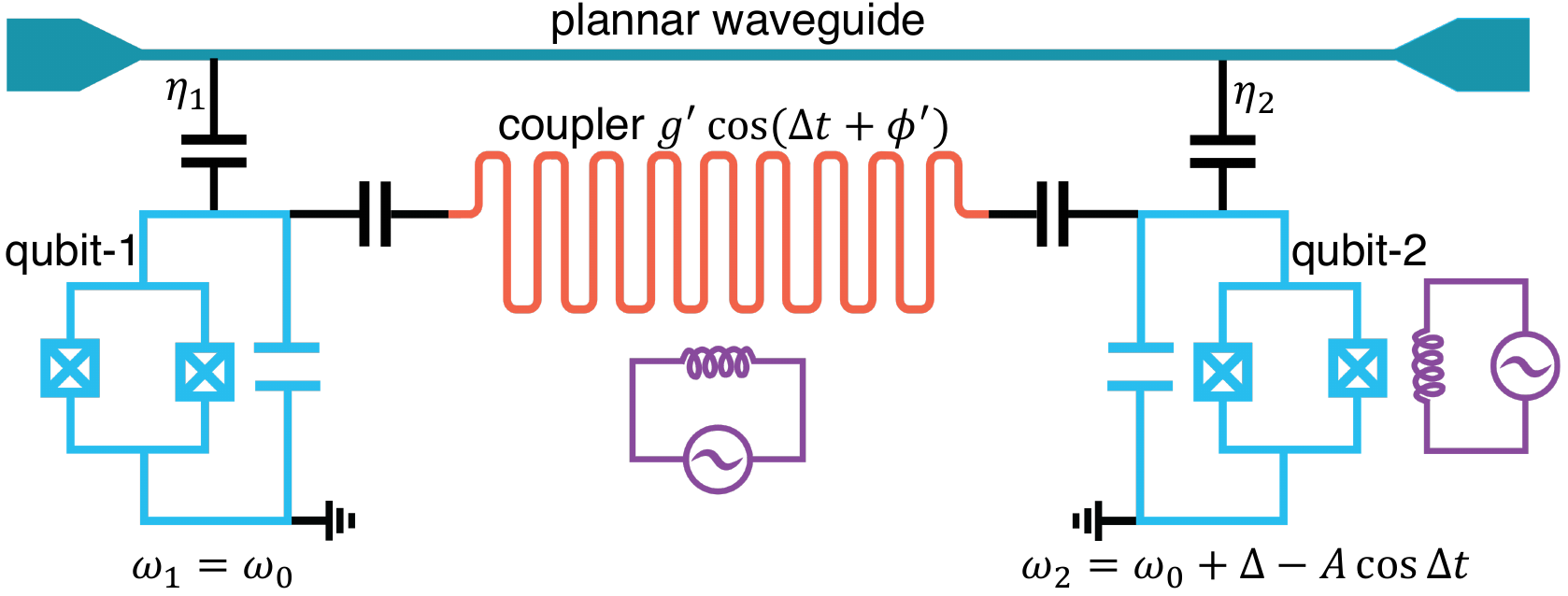}
\centering
\caption{\label{fig:5} Schematic of the experimental implementation. The frequency of qubit-1 is $\omega_0$. The frequency $\omega_2 = \omega_0+\Delta-A\cos\Delta t$ of qubit-2 and the qubit-qubit interaction $g^{\prime}\cos(\Delta+\phi')$ 
are modulated by two microwaves. The two qubits are coupled to the waveguide with strengths $\eta_1$ and $\eta_2$.}
\vspace{-0.5cm}
\end{figure}

\section{Implementation and Conclusion\label{sec6}}
The circuit QED system could be a promising platform to realize the symmetry-breaking atom-atom interaction $\hat{H}_a$ in Eq.~(\ref{eq:atom_interaction}) with the continuously adjustable phase $\phi$~\cite{Kannan2023Ondemand,Joshi2023Resonance}. As illustrated in Fig.~\ref{fig:5}, two transmon qubits (artificial atoms) are coupled to a coplanar waveguide. The typical resonant transition frequency of the qubits is a few gigahertz, such as $\omega_0/2\pi=4.9$~GHz~\cite{Kannan2023Ondemand}. The detuning $\Delta$ of the two qubits can be tuned via flux bias or Josephson inductance~\cite{krantz2019quantum,ripoll2020quantum}. A tunable coupler is employed to achieve the direct interaction between the two qubits~\cite{roushan2017chiral}. The frequency of qubit-2 and the qubit-qubit coupling are modulated by two microwaves with identical frequency $\Delta/2\pi = 200$~MHz and tunable phase difference $\phi^{\prime}$~\cite{roushan2017chiral,McKay2016Universal,Wu2018}. Under the condition $\Delta\gg A,\, g'$, our proposed model Hamiltonian can be realized with $\eta_1 = \eta_2 J_1(A/\Delta)=\eta$ and $g^{\prime}\left[J_{0}(A/\Delta)e^{i\phi^{\prime}}+J_{2}(A/\Delta)e^{-i\phi^{\prime}}\right]/2\equiv g\exp(i\phi)$ (see Appendix~\ref{Appendix6}), where $J_n(x)$ is the $n$-th Bessel function of the first kind and the phase $\phi$ can be tuned on demand.
The experimentally accessible parameters
are chosen as $g/2\pi=\gamma/2\pi=3.2$MHz~\cite{Kannan2023Ondemand}.

We introduce an innovative QBS composed of a Hermitian $\mathcal{PT}$-symmetric system. Our QBS enables fine control of the reflection and transmission coefficients, as well as the quantum coherence of output photons~\cite{tomm2023photon,masters2023simultaneous,Jeannic2021experimental}. To achieve a four-port BS with distinct input and output ports, we can replace the two-way waveguide with two separate unidirectional waveguides~\cite{Jan2014chira,Scarpelli2019beta}. The basic concepts underlying this study can be extended to multi-atom systems, enabling manipulations of higher-order photon coherence. Our results could open up new avenues for leveraging asymmetric higher-order quantum coherence in the development of non-reciprocal quantum devices~\cite{PRL2023nonreciprocal,Yang2020Inverse,Otterstrom2021Nonreciprocal,zhou2008quantum,dayan2008photon}.

\section*{ACKNOWLEDGMENTS}
L.P.Y. is supported by the Innovation Program for Quantum Science and Technology (No. 2023ZD0300700) and the National Natural Science Foundation of China (NSFC) Grant No. 12275048. Y.C. is supported by the NSFC Grants No. U2141237 and No.62101070.

\appendix

\section{Symmetry analysis of the system\label{Appendix1}}
In this section, we introduce the model of our system. The Hamiltonian of our quantum beam splitter contains three parts $\hat{H}=\hat{H}_{a}+\hat{H}_{p}+\hat{H}_{\rm int}$.
The Hamiltonian 
\begin{equation}
\hat{H}_{a}=\omega_0\left(\hat{\sigma}_{1}^{\dagger}\hat{\sigma}_{1}+\hat{\sigma}_{2}^{\dagger}\hat{\sigma}_{2}\right)+ge^{i\phi}\hat{\sigma}_{1}^{\dagger}\hat{\sigma}_{2}+ge^{-i\phi}\hat{\sigma}_{1}\hat{\sigma}_{2}^{\dagger}, \label{eq:H_a}
\end{equation}
describes the two atoms with direct coupling and $\sigma_{i}=\left|g_{i}\right\rangle \left\langle e_{i}\right|$. The phase $\phi$ in the coupling is a tunable parameter.
The Hamiltonian of the waveguide photons is given by
\begin{equation}
\hat{H}_{p}=\int dk\sum_{\lambda=r,l}\omega_{k,\lambda}\hat{b}_{k,\lambda}^{\dagger}\hat{b}_{k,\lambda}, \label{eq:H_p}
\end{equation}
where the subscripts $\lambda=r,l$ denote the left- and right-propagating
modes, respectively. The frequencies of the waveguide
modes can be expanded around atomic transition frequency as $\omega_{k,\lambda}\approx\omega_0+\lambda v_{g}k=v_{g}(k_{0}+\lambda k)$,
where $v_{g}=\omega_0/k_0$ is the group velocity of the photons at the center wavenumber $k_0$.
For convenience, we take $v_{g}=1$ in the following. The coupling
between the atoms and photons follows the Jaynes-Cumming form
\begin{align}
\hat{H}_{\rm int} =\sum_{i=1,2}\eta\int dk\hat{\sigma}_{i}\left[\hat{b}_{k,r}^{\dagger}e^{-i( k_{0}+ k)x_{i}}+\hat{b}_{k,l}^{\dagger}e^{i( k_{0}- k)x_{i}}\right]+{\rm H.c.}, \label{eq:Hint}
\end{align}
where $\eta$ denotes the coupling strength. In the following, we will
let $x_{1}=-d/2$, $x_{2}=d/2$, and $d$ is the separation of the two atoms. In the following, we perform the calculation in the rotating frame with respect to $\omega_0 \left(\hat{\sigma}_1^{\dagger}\hat{\sigma}_1+\hat{\sigma}_2^{\dagger}\hat{\sigma}_2+\int dk \sum_{\lambda}\hat{b}_{k,\lambda}^{\dagger}\hat{b}_{k,\lambda}\right)$. The free Hamiltonian of the atoms and photons changes to
\begin{equation}
 \hat{H}_0\! =\! \hat{H}_a + \hat{H}_p \!=\!  ge^{i\phi}\hat{\sigma}_{1}^{\dagger}\hat{\sigma}_{2}\!+\!ge^{-i\phi}\hat{\sigma}_{2}^{\dagger}\hat{\sigma}_{1}\!+\!\int \!\!dk \!\sum_{\lambda} \!\lambda k \hat{b}_{k,\lambda}^{\dagger}\hat{b}_{k,\lambda}.\!  
\end{equation}

In this section, we give the symmetry analysis of the system and reveal the symmetry relations of the photon scattering amplitudes. Under the space inversion operation $\hat{\mathcal{P}}$, then we have the following relations for the photonic operators
\begin{equation}
\hat{\mathcal{P}}^{-1} \hat{b}_{k,l}\hat{\mathcal{P}} = \hat{b}_{-k,r},\ \hat{\mathcal{P}}^{-1} \hat{b}_{k,r}\hat{\mathcal{P}} = \hat{b}_{-k,l}.  \label{eq:parityinv1}
\end{equation}
We also need to perform the transformation for the coordinates $x\rightarrow -x$, and interchange the indices $1\leftrightarrow 2$ of the atoms. We can verify that apart from the direct coupling term between the two atoms, all the other terms remain invariant under space inversion. Consequently, the overall system's parity symmetry is broken and we obtain the following relation
\begin{equation}
\hat{\mathcal{P}}^{-1}\hat{H}\left( \phi \right) \hat{\mathcal{P}}=\hat{H}\left( -\phi \right). \label{eq:parityinv2}
\end{equation}
Similarly, under the time reversal operation $\hat{\mathcal{T}}$, we have
\begin{equation}
\hat{\mathcal{T}}^{-1} \hat{b}_{k,l}\hat{\mathcal{T}} = \hat{b}_{-k,r},\ \hat{\mathcal{T}}^{-1} \hat{b}_{k,r}\hat{\mathcal{T}} = \hat{b}_{-k,l}.  
\end{equation} 
We also need to take the complex conjugate of the Hamiltonian, i.e., $\hat{H}\rightarrow \hat{H}^{*}$. We can confirm that the direct atomic coupling term also breaks the time-reversal symmetry
\begin{equation}
\hat{\mathcal{T}}^{-1 }\hat{H}\left( \phi \right) \hat{\mathcal{T}}=\hat{H}\left( -\phi \right).
\end{equation}

Although both parity and time-reversal symmetries have been broken, the parity-time ($\mathcal{PT}$) symmetry of the system is preserved
\begin{equation}
\hat{\mathcal{T}}^{-1}\hat{\mathcal{P}}^{-1}\hat{H}\left( \phi \right) \hat{\mathcal{P}}\hat{\mathcal{T}}=\hat{H}\left( \phi \right).
\end{equation}
Through symmetry analysis, we can directly derive useful scattering identities, such as
\begin{align}
 \mathcal{A}_{r\rightarrow l}\left( -p,k\right) & = \left\langle
gg\right\vert \left\langle 0\right\vert \hat{b}_{-p,l}\hat{S}\hat{b}
_{k,r}^{\dagger
}\left\vert 0\right\rangle \left\vert gg\right\rangle\nonumber\\
& = \left\langle
gg\right\vert \left\langle 0\right\vert \hat{b}_{k,r}\hat{S}\hat{b}
_{-p,l}^{\dagger
}\left\vert 0\right\rangle \left\vert gg\right\rangle = \mathcal{A}_{l\rightarrow r}\left( k,-p\right),
\end{align} 
where $\left\vert 0\right\rangle$ is the vacuum state of the photons and $\hat{S}=\exp\left(i\hat{H}_0t_f\right)\exp\left(i\hat{H}T\right)\exp\left(-i\hat{H}_0t_i\right)$ is the scattering matrix  (see Sec.~\ref{sec:scatteringchannels}). Energy conservation law brings a delta function $\delta (k-p)$ in these two
scattering amplitudes. Hence, at the single-photon level, the reflection
amplitudes for right-moving and left-moving incident photons are the same.
Consequently, the transmission probabilities for these two situations are
also the same. Namely, the single-photon scattering process is reciprocal.
However, this does not hold for multi-photon scattering. For instance, the $%
\mathcal{PT}$-symmetry gives the following equality for the two-photon
reflection amplitudes
\begin{align}
\left\langle gg\right\vert \left\langle 0\right\vert
\hat{b}_{-p_{2},l}&\hat{b}_{-p_{1},l}\hat{S}\hat{b}_{k_{2},r}^{\dagger }\hat{b}_{k_{1},r}^{\dagger
}\left\vert 0\right\rangle \left\vert gg\right\rangle \nonumber\\
& = \left\langle
gg\right\vert \left\langle 0\right\vert
\hat{b}_{k_{2},r}\hat{b}_{k_{1},r}\hat{S}\hat{b}_{-p_{2},l}^{\dagger }b_{-p_{1},l}^{\dagger
}\left\vert 0\right\rangle \left\vert gg\right\rangle
\end{align}%
along with the energy conservation condition $\delta (k_{1}+k_{2}-p_{1}-p_{2})$. However, this condition does not guarantee that the two-photon reflection process is reciprocal, i.e., $\left\{ k_{1},k_{2}\right\} =\left\{
p_{1},p_{2}\right\} $. This nonreciprocity is not only
illustrated in the reflection probability, but also in the higher-order
correlation functions.

Using the parity inversion relations (\ref{eq:parityinv1}) and (\ref{eq:parityinv2}), we can also obtain very useful relations for our system, such as
\begin{align}
\left\langle gg\right\vert \left\langle 0\right\vert &\hat{b}_{k_{f},\lambda _{f}}\hat{S}\left( \phi \right) \hat{b}_{k_{i},\lambda _{i}}^{\dagger }\left\vert 0\right\rangle \left\vert
gg\right\rangle \nonumber\\
& = \left\langle gg\right\vert \left\langle 0\right\vert \hat{\mathcal{P}}^{-1}\hat{b}_{k_{f},\lambda _{f}}%
\hat{\mathcal{P}} \hat{S}\left( -\phi \right) 
\hat{\mathcal{P}}^{-1}\hat{b}_{k_{i},\lambda _{i}}^{\dagger }\hat{\mathcal{P}}
\left\vert 0\right\rangle \left\vert gg\right\rangle.
\end{align}
This identity can be readily extended to multi-photon scattering scenarios. Such scattering relations can significantly simplify calculations. For instance, the transmission amplitude for a single incident right-moving photon can be obtained by substituting $\phi$ with $-\phi$ in the transmission amplitude for a left-moving incident photon:
\begin{equation*}
\left\langle gg\right\vert \left\langle 0\right\vert \hat{b}_{k,r}\hat{S}\left( \phi
\right) \hat{b}_{p,r}^{\dagger }\left\vert 0\right\rangle \left\vert
gg\right\rangle\! =\!\left\langle gg\right\vert \left\langle 0\right\vert
\hat{b}_{-k,l}\hat{S}\left( -\phi \right) \hat{b}_{-p,l}^{\dagger }\left\vert
0\right\rangle \left\vert gg\right\rangle .
\end{equation*}

\section{Input-Output relations and quantum regression theorem\label{Appendix2}}
In this section, we introduce the theoretical tools employed in the main text and outline a numerical approach for calculating photonic correlation functions

\subsection{Input-output relation}
The Heisenberg equations for the ladder operators 
 of atoms and photons are
given by
\begin{align}
\frac{d}{dt}\hat{b}_{k,\lambda}(t) & =-i\lambda k\hat{b}_{k,\lambda}(t)-i\eta\sum_{i}\hat{\sigma}_{i}(t)e^{-i(\lambda k_{0}+k)x_i},\\
\frac{d}{dt}\hat{\sigma}_{i}(t) & =i[\hat{H}_{a},\hat{\sigma}_{i}(t)]-i\eta\sum_{\lambda}\int dk\hat{b}_{k,\lambda}(t)e^{i(\lambda k_{0}+k)x_i}.\label{eq:ai}
\end{align}
To derive the standard input-output relations, we give the formal solution
of $\hat{b}_{k,\lambda}(t)$ with both initial conditions and final output conditions
\begin{align}
\hat{b}_{k,\lambda}(t) &\! =\! \hat{b}_{k,\lambda}(t_{i})e^{-i\lambda k(t-t_{i})}\!-\!i\eta\!\sum_{i}\!\int_{t_{i}}^{t}\!\!dt'\hat{\sigma}_{i}(t')e^{-i\lambda k(t-t')}e^{-i(\lambda k_{0}+k)x_i},\label{eq:bk1} \\
\hat{b}_{k,\lambda}(t) &\! =\! \hat{b}_{k,\lambda}(t_{f})e^{-i\lambda k(t-t_{f})}\!+\!i\eta\!\sum_{i}\!\int_{t}^{t_{f}}\!\!\!dt'\hat{\sigma}_{i}(t')e^{-i\lambda k(t-t')}e^{-i(\lambda k_{0}+k)x_i},\label{eq:bk2}
\end{align}
with $t_{i} < t <t_f$. Subtracting Eq.~(\ref{eq:bk2}) from Eq.~(\ref{eq:bk1}) and integrating over $k$, we obtain the input-output relation under the Markov approximation
\begin{align}
&\!\!\!\!\!\!\hat{b}_{{\rm out},\lambda}(t) \!=\!  \hat{b}_{{\rm in},\lambda}(t)\!-\!\frac{i\eta}{\sqrt{2\pi}}\!\int \!\!dk\!\int_{t_{i}}^{t_{f}}\!\!\!dt'\!\!\sum_{i}\hat{\sigma}_{i}(t')e^{-i\lambda k(t-t')}e^{-i(\lambda k_{0}+k)x_i} \nonumber\\
\!\!\approx &\hat{b}_{{\rm in},\lambda}(t)\!-\!i\!\sqrt{\Gamma}\!\sum_{i}\hat{\sigma}_{i}(t)\Theta(t\!+\!\lambda x_i\!-\!t_{i})\Theta(t_{f}\!-\!t\!-\!\lambda x_i)e^{-i\lambda k_{0} x_i}\!,\label{eq:input-output}\!\!
\end{align}    
where $\Gamma=2\pi \eta^{2}$ and the step functions come from the fact $t_i \leq t'=t+\lambda x_i\leq t_{f}$. The input and output operators are defined as
\begin{align}
\hat{b}_{{\rm in},\lambda}(t) & =\frac{1}{\sqrt{2\pi}}\int dke^{-i\lambda k(t-t_{i})}\hat{b}_{k,\lambda}(t_{i}),\\
\hat{b}_{{\rm out},\lambda}(t) & = \frac{1}{\sqrt{2\pi}}\int dke^{-i\lambda k(t-t_{f})}\hat{b}_{k,\lambda}(t_{f}).
\end{align}
In the following, we focus solely on the statistical properties of the far fields. Consequently, we take the limits of $t_{i}\rightarrow-\infty$ and $t_{f}\rightarrow\infty$, while omitting the step functions in Eq.~(\ref{eq:input-output}).

We can readily verify that the defined input and output operators satisfy the following commutation relation
\begin{align}
\left[\hat{b}_{{\rm in},\lambda}(t),\hat{b}_{{\rm in},\lambda'}^{\dagger}(t')\right] & =\delta_{\lambda\lambda'}\delta(t-t'),\\
\left[\hat{b}_{{\rm out},\lambda}(t),\hat{b}_{{\rm out},\lambda'}^{\dagger}(t')\right] & =\delta_{\lambda\lambda'}\delta(t-t').
\end{align}
Due to the causality relation, we have the following commutation relations between an arbitrary atomic operator $\hat{O}$ and the input and output fields~\cite{carmichael2002statistical,gardiner2004quantum}
\begin{align}
[\hat{O}(t),\hat{b}_{{\rm in},\lambda}(t')] &  = 0,  {\rm for}\ t'>t \label{eq:causality1} \\
[\hat{O}(t),\hat{b}_{{\rm out},\lambda}(t')] & = 0, {\rm for}\ t'<t. \label{eq:causality2}
\end{align}

\subsection{Initial state of the waveguide modes under coherent drivings}
In the following, we will examine the statistical properties of the output photons from our beam splitter when subjected to two weak coherent-state inputs. In experiments, two extremely long pulses will be utilized. In this scenario, the initial state of the waveguide photons can be described by
$|\Psi(t_i)\rangle = |\alpha_l\rangle\otimes |\alpha_r\rangle$, where
\begin{equation}
|\alpha_\lambda\rangle = e^{-|\alpha_{\lambda}|^2} e^{\alpha_{\lambda}\hat{B}_{\lambda}^{\dagger}}|0\rangle. \label{eq:coherentstate} 
\end{equation}
Here, we have introduced a single-photon wave-packet creation operator 
\begin{equation}
\hat{B}^{\dagger}_{\lambda} \equiv \int dk \xi_{\lambda} (k)\hat{b}^{\dagger}_{k,\lambda}, 
\end{equation}
with normalized amplitude $\int dk |\xi_{\lambda} (k)|^2 =1$ and commutation relation $[\hat{B}_{\lambda},\hat{B}_{\lambda}^{\dagger}]=1$. We have the following relation
\begin{equation}
\hat{b}_{k,\lambda}|\alpha_{\lambda}\rangle = \alpha_{\lambda}\xi (k)|\alpha_{\lambda}\rangle.   
\end{equation}

In cases involving near-continuous-wave laser drivings in the modes $b_{p,r}$ and $b_{-p,l}$, the corresponding amplitude function approaches a delta function $\xi_{\lambda}(k)\propto \delta (k-\lambda p)$. For the sake of simplicity, we employ an extremely narrow Gaussian function to model the amplitude
\begin{equation}
 \xi_{\lambda} (k) = \left(\frac{1}{2\pi\sigma^2}\right)^{1/4}\exp \left[-\frac{(k-\lambda p)^2}{4\sigma^2}\right],   
\end{equation}
where $\sigma$ is a small positive number. Then, the single-photon wave-packet creation $
\hat{B}^{\dagger}_{\lambda} \approx \sqrt{\epsilon}\hat{b}^{\dagger}_{\lambda p,\lambda}$,  with factor $\epsilon=2\sqrt{2\pi}\sigma$. The corresponding wave packet functions of two driving pulses 
\begin{align}
 \tilde{\xi}(x) &= \frac{1}{\sqrt{2\pi}}\int dk e^{i\lambda(k_0+\lambda k) x} \xi_{\lambda}(k)\nonumber\\ 
 & = \sqrt{\epsilon} \exp{\left[-\sigma^2x^2+i\lambda (k_0 +p)x\right]},
\end{align}
could be approximated as plane waves with amplitude $\sqrt{\epsilon}$. 

\subsection{Master equation for atoms and quantum regression theorem}
To study the statistics of the output photons, we need to evaluate their wave function and higher-order correlation functions. By utilizing the input-output relations and the commutation relations mentioned in the previous subsection, the multi-time photon correlation functions can be transformed into atomic correlations. The atomic correlations can be evaluated using the master equation and the quantum regression theorem~\cite{Shi2015multiphoton,Chang2016deterministic,Caneva2015Quantum}.

In addition to the Heisenberg equations, the dynamics of the atoms can also be described by the standard quantum master equation for coherent-state input cases. The density matrix of the whole system satisfies the Liouville–von Neumann equation
\begin{equation}
\partial_{t}\rho(t)=\partial_{t}\left[e^{-i\hat{H}t}\rho(0)e^{i\hat{H}t}\right]=-i\left[\hat{H},\rho(t)\right].
\end{equation}
To obtain a standard master equation for the atoms, we can take the initial of all the waveguide modes as the vacuum state and treat the coherent-state inputs as classical drivings~\cite{Shi2015multiphoton,zhang2023limits,gheri1998photon}. By eliminating the waveguide modes, we obtain the master equation of the atoms with drivings
\begin{align}
&\partial_{t}\rho_{a}(t)\equiv\mathcal{L}\rho_{a}(t)\nonumber\\
&=\!-i\left[\hat{H}_{a}^{\prime},\rho_{a}\right]+\sum_{i,j}\Gamma_{ij}\left[2\hat{\sigma}_{i}\rho_{a}\hat{\sigma}_{j}^{\dagger}\!-\!\rho_{a}\hat{\sigma}_{i}^{\dagger}\hat{\sigma}_{j}\!-\!\hat{\sigma}_{i}^{\dagger}\hat{\sigma}_{j}\rho_{a}\right].\label{eq:mastereq}
\end{align}
The waveguide modes lead to the decoherence of the atoms with decay rates 
\begin{equation}
 \Gamma_{ij}=\Gamma\cos k_0(x_i - x_j). \label{eq:Gammaij}  
\end{equation}
The atom-photon interaction also induces a coherent dipole-dipole interaction between the two atoms. As a result, the Hamiltonian of the atoms changes to
\begin{equation}
\hat{H}_{a}^{\prime}=g_{{\rm eff}}\hat{\sigma}_{1}^{\dagger}\hat{\sigma}_{2}+g_{{\rm eff}}^{*}\hat{\sigma}_{1}\hat{\sigma}_{2}^{\dagger} + \hat{H}_{\rm drive},
\end{equation}
with effective coupling strength
\begin{equation}
 g_{{\rm eff}}=ge^{i\phi}+\Gamma\sin\theta,\label{eq:geff} 
\end{equation}
and $\theta = k_{0}|x_{1}-x_{2}|=k_0 d$.
The coherent-state pumpings from the two sides of the waveguide are effectively described by adding coherent driving terms to the atomic Hamiltonian
\begin{align}
\hat{H}_{\rm drive} = & \Omega_r \sum_{i} \hat{\sigma}_{i}^{\dagger}e^{i(k_0+p_r)x_i}e^{-ip_r(t-t_i)}\nonumber \\
& + \Omega_l \sum_{i} \hat{\sigma}_{i}^{\dagger}e^{-i(k_0-p_l)x_i}e^{ip_l(t-t_i)} + {\rm h.c.},\label{eq:Hdrive}
\end{align}
where $k_0 +\lambda p_\lambda$ are the wavenumbers of the input modes and $\Omega_{\lambda}=\sqrt{\epsilon\Gamma}\alpha_{\lambda}$ are the  pumping amplitudes. In the following, we focus more on the case with $p_l = p_r=p$, i.e., the interference of two beams with the same frequency.

Similar to the density matrix $\rho_a (t)$ of the atoms, the dynamics of an arbitrary operator $\hat{Q}(t)$
satisfying the von Newmann-Liouville equation
$\partial_{t}\hat{Q}(t)=-i\left[\hat{H},\hat{Q}(t)\right]$
could also be described by an effective master equation
\begin{equation}
\partial_{t}{\rm Tr}_{B}\hat{Q}(t)=\mathcal{L}{\rm Tr}_{B}\hat{Q}(t),
\end{equation}
or
\begin{equation}
{\rm Tr}_{B}\hat{Q}(t)=e^{\mathcal{L}t}{\rm Tr}_{B}\hat{Q}(0).
\end{equation}
Then, we can obtain the quantum regression theorem for arbitrary system
operators $\hat{O}_{i}$ ordered by time~\cite{carmichael2002statistical}. We take a three-time correlation function
with $t_{3}>t_{2}>t_{1}$ for example
\begin{align}
 & {\rm Tr}\left[\hat{O}_{3}(t_{3})\hat{O}_{2}(t_{2})\hat{O}_{1}(t_{1})\rho(0)\right]\nonumber \\
& = {\rm Tr}_{a}\left\{\hat{O}_{3}e^{\mathcal{L}(t_{3}-t_{2})}\hat{O}_{2}e^{\mathcal{L}(t_{2}-t_{1})}\hat{O}_{1}e^{\mathcal{L}t_{1}}{\rm Tr}_{B}\rho(0)\right\}.
\end{align}

\subsection{Numerical method for photonic correlation functions\label{sec:numberical}}
All the relevant properties of the output photons can be obtained from multi-time correlation functions, which can be numerically evaluated by combining the input-output relation and the quantum regression theorem. Here, we illustrate the key idea by considering examples of the first- and second-order correlation functions with delay $\tau\geq 0$,\begin{widetext}
\begin{align}
 G^{(1)}_{\lambda}(\tau) & = \langle gg|\otimes\langle \Psi (t_i)|\hat{\psi}^{\dagger}_{\lambda}(X,t_f) \hat{\psi}_{\lambda}(X+\lambda\tau,t_f)  |\Psi (t_i)\rangle\otimes|gg\rangle, \\
G^{(2)}_{\lambda}(\tau) & = \langle gg|\otimes\langle \Psi (t_i)|\hat{\psi}^{\dagger}_{\lambda}(X,t_f)\hat{\psi}^{\dagger}_{\lambda}(X+\lambda\tau,t_f) \hat{\psi}_{\lambda'}(X+\lambda\tau,t_f) \hat{\psi}_{\lambda}(X,t_f)  |\Psi (t_i)\rangle\otimes|gg\rangle,
\end{align}    
\end{widetext}
where we only consider the correlation of photons at the same output port and 
\begin{equation}
   \hat{\psi}_\lambda (X,t_f)= \frac{1}{\sqrt{2\pi}}\int d k b_{k,\lambda} (t_f)e^{i\lambda(k_0+\lambda k)X}, 
\end{equation}
is the field operator of the output photon in the Heisenberg picture and 
$|\Psi (t_i)\rangle$ is the initial state of the waveguide photons including the coherent-state inputs. 

The presented method can be readily applied to more general cases. Utilizing the input-output relation, we have\begin{widetext}
\begin{align}
G_{\lambda}^{(1)}(\tau) 
= &\frac{1}{2\pi}\int dke^{-i\lambda(k_{0}+\lambda k)X}\int dpe^{i\lambda(k_{0}+\lambda p)(X+\tau)}\langle gg|\otimes\langle\Psi(t_{i})|\hat{b}_{k,\lambda}^{\dagger}(t_{f})\hat{b}_{p,\lambda}(t_{f})|\Psi(t_{i})\rangle\otimes|gg\rangle\\
 = & e^{i\lambda k_{0}\tau}\langle gg|\otimes\langle\Psi(t_{i})|\hat{b}_{{\rm out},\lambda}^{\dagger}(t_{f}-\lambda X)\hat{b}_{{\rm out},\lambda}(t_{f}-\lambda X-\tau)|\Psi(t_{i})\rangle\otimes|gg\rangle\\
 \approx & e^{i\lambda k_{0}\tau}\langle gg|\otimes\langle\Psi(t_{i})|\left[\sqrt{\epsilon}\alpha_{\lambda}^{*}e^{i\lambda p(T-\lambda X)}+i\Gamma\sum_{i}\hat{\sigma}_{i}^{\dagger}(t_f-\lambda X)e^{i\lambda k_{0}x_{i}}\right]\nonumber\\
 & \times \left[\sqrt{\epsilon}\alpha_{\lambda}e^{-i\lambda p(T-\lambda X-\tau)}-i\Gamma\sum_{j}\hat{\sigma}_{j}(t_f-\lambda X-\tau)e^{-i\lambda k_{0}x_{j}}\right]|\Psi(t_{i})\rangle\otimes|gg\rangle. 
\end{align}    
\end{widetext}
where $T=t_f -t_i$ and the step function in the third step has been omitted for the
far fields with $\lambda X\gg|x_{i}|$. The mean value of the atomic
operators can be evaluated via the master equation (\ref{eq:mastereq}) and the quantum
regression theorem numerically, e.g.,\begin{widetext}
\begin{align}
\langle gg|\otimes\langle\Psi(t_{i})|\hat{\sigma}_{i}^{\dagger}(t_f-\lambda X)|\Psi(t_{i})\rangle\otimes|gg\rangle & ={\rm Tr}_{a}\left[\hat{\sigma}_{i}^{\dagger}e^{\mathcal{L}(T-\lambda X)}\left|gg\right\rangle \left\langle gg\right|\right],\\
\langle gg|\otimes\langle\Psi(t_{i})|\hat{\sigma}_{i}^{\dagger}(t_f-\lambda X)\hat{\sigma}_{j}^{\dagger}(t_f-\lambda X-\tau)|\Psi(t_{i})\rangle\otimes|gg\rangle & ={\rm Tr}_{a}\left[\hat{\sigma}_{i}^{\dagger}e^{\mathcal{L}\tau}\hat{\sigma}_{j}e^{\mathcal{L}(T-\lambda X)}\left|gg\right\rangle \left\langle gg\right|\right].
\end{align}    
\end{widetext}

For $\tau = 0$, the first-order correlation function gives the mean photon number density of $\lambda$-output port
\begin{equation}
G^{(1)}_{\lambda}(0)
\!=\!{\rm Tr}_a\!\left[\hat{\mathscr{O}}_{\lambda}(T\!-\!\lambda X)e^{\mathcal{L}(T\!-\!\lambda X)}|gg\rangle\langle gg|\hat{\mathscr{O}}_{\lambda}^{\dagger}(T\!-\!\lambda X)\right],  
\end{equation}
where we have introduced an output operator for atoms
\begin{equation}
\hat{\mathscr{O}}_{\lambda} (t) \equiv  \sqrt{\epsilon}\alpha_{\lambda}e^{-i\lambda p t}-i\Gamma\sum_{i}e^{-i\lambda k_{0}x_{i}}\hat{\sigma}_{i}.  
\end{equation}
To simplify the simulation, we can eliminate the oscillating phase factor $\exp(-i\lambda p t)$ in the operator $\hat{\mathscr{O}}_{\lambda} (t)$ and the driving Hamiltonian (\ref{eq:Hdrive}) by selecting an appropriate rotating frame as shown in the main text. To obtain the reflectance and transmittance of right-moving photons incident from the left port, we can let $\alpha_{l}=0$ and re-scale the output photon intensity with $\epsilon |\alpha_r|^2$:
\begin{equation}
 |r_{r\rightarrow l}(p)|^2 = \frac{G^{(1)}_{l}(0)}{\epsilon |\alpha_r|^2},\ |t_{r\rightarrow r}(p)|^2 = \frac{G^{(1)}_{r}(0)}{\epsilon |\alpha_r|^2}    
\end{equation}
where $r_{r\rightarrow l}(p)$ and $t_{r\rightarrow r}(p)$ are the the reflection and transmission coefficients, respectively. The reflectance and transmittance of left-moving photons incident from the right port can be obtained similarly.

Similarly, the second-order correlation function of the output photon in the case of weak coherent-state drivings can be approximated as\begin{widetext}
\begin{align}
G_{\lambda}^{(2)}(\tau) 
= & \langle gg|\otimes\langle\Psi(t_{i})|\hat{b}_{{\rm out},\lambda}^{\dagger}(t_{f}-\lambda X)\hat{b}_{{\rm out},\lambda}^{\dagger}(t_{f}-\lambda X-\tau)\hat{b}_{{\rm out},\lambda}(t_{f}-\lambda X-\tau)\hat{b}_{{\rm out},\lambda}(t_{f}-\lambda X)|\Psi(t_{i})\rangle\otimes|gg\rangle
\\
= & {\rm Tr}\left[\hat{b}_{{\rm out},\lambda}(t_{f}-\lambda X)\hat{b}_{{\rm out},\lambda}(t_{f}-\lambda X-\tau)|\Psi(t_{i})\rangle\otimes|gg\rangle\langle gg|\otimes\langle\Psi(t_{i})|\hat{b}_{{\rm out},\lambda}^{\dagger}(t_{f}-\lambda X-\tau)\hat{b}_{{\rm out},\lambda}^{\dagger}(t_{f}-\lambda X)\right]\\
= & {\rm Tr}_a\left\{\hat{\mathscr{O}}_{\lambda}(T-\lambda X)e^{\mathcal{L}\tau}\left[\hat{\mathscr{O}}_{\lambda}(T-\lambda X-\tau)e^{\mathcal{L}(T-\lambda X-\tau)}\left(|gg\rangle\langle gg|\right)\hat{\mathscr{O}}_{\lambda}^{\dagger}(T-\lambda X-\tau)\right]\hat{\mathscr{O}}_{\lambda}^{\dagger}(T-\lambda X)\right\},
\end{align}    
\end{widetext}
where we re-order the output operators according to time in the second step. The coherence function $g^{(2)}(\tau)$ can be obtained by re-scaling the correlation function $G^{(2)}_{\lambda}(\tau)$ with its steady value $G^{(2)}_{\lambda}(\infty)$, which is equal to the square of the steady value of $G^{(1)}(0)$. Next, we show how to obtain the properties of the output photon via the scattering method.
 
\section{Scattering Method for Determining Reflection and Transmission Coefficients\label{Appendix3}} 

In the case of weak input, the mean photon numbers at the two output ports of the beam splitter are determined by the reflectance and transmittance, which are predominantly determined by the single-photon scattering processes in the weak driving case. In the following sections, we will employ the scattering method to derive analytical expressions for the reflection and transmission coefficients. 

\subsection{Scattering coefficients for a right-moving single-photon input}
We consider the scenario where a right-moving plane-wave single photon with a wavenumber $k_0+p$ is incident from the left side of the waveguide. The scattering amplitude in the left-moving mode with a wavenumber $-k_0-k$ is given by:\begin{widetext}
\begin{align}
\mathcal{A}_{r\rightarrow l}(-k,p) = & \langle gg|\otimes \langle 0|\hat{b}_{-k,l}e^{i\hat{H}_0 t_f}e^{-i\hat{H}T}e^{-i\hat{H}_0 t_i}\hat{b}_{p,r}^{\dagger}|0\rangle \otimes |gg\rangle  \\
= & \frac{1}{2\pi}e^{i(kt_f-pt_i)}\int_{t_i}^{t_f} dt_{2}e^{ik(t_{2}-t_{f})}\int_{t_i}^{t_f} dt_{1}e^{-ip(t_{1}-t_{i})}\langle gg|\otimes \langle 0|\hat{b}_{{\rm out},l}(t_{2})\hat{b}_{{\rm in},r}^{\dagger}(t_{1})|0\rangle \otimes |gg\rangle \\
= & \frac{-\Gamma}{2\pi}e^{i(kt_f-pt_i)}\sum_{ij}e^{ik_{0}(x_{i}+x_{j})}\int_{t_i}^{t_f} dt_{2}e^{ik(t_{2}-t_{f})}\int_{t_i}^{t_{2}}dt_{1}e^{-ip(t_{1}-t_{i})}\langle gg|\otimes \langle 0|\hat{\sigma}_{i}(t_{2})\hat{\sigma}_{j}^{\dagger}(t_{1})|0\rangle \otimes |gg\rangle \label{eq:correlator1}\\
= & \frac{-\Gamma}{2\pi}e^{i(kt_f-pt_i)}\sum_{ij}e^{ik_{0}(x_{i}+x_{j})}\int_{t_i}^{t_f} dt_{2}e^{ik(t_{2}-t_{f})}\int_{t_i}^{t_{2}}dt_{1}e^{-ip(t_{1}-t_{i})}{\rm Tr}_a\left[\hat{\sigma}_{i}e^{\mathcal{L}(t_{2}-t_{1})}\hat{\sigma}_{j}^{\dagger}|gg\rangle\langle gg|\right],
\end{align}    
\end{widetext}
where $T=t_f - t_i$ is the evolution time. We have used the input-output relation and the commutation relations (\ref{eq:causality1}) and (\ref{eq:causality2}) in the third step and the fact $e^{\mathcal{L}t}|gg\rangle\langle gg|=|gg\rangle\langle gg|$ in the last step.

The time-evolution of the atomic operators $\hat{\sigma}^{\dagger}_j |gg\rangle\langle gg|$ can be evaluated using the master equation (\ref{eq:mastereq}). It is important to note that, unlike the numerical method in section~\ref{sec:numberical}, the pumping effect has already been considered in the input operator, and the photonic state is the vacuum state for the atomic correlation function in Eq.~(\ref{eq:correlator1}). Consequently, we need to set the pumping strength $\alpha_{\lambda}$ in the driving Hamiltonian (\ref{eq:Hdrive}) to zero. Moreover, it can be verified that the jumping terms $2\hat{\sigma}_i \rho_a \hat{\sigma}^{\dagger}_j$ in the master equation do not contribute to the scattering amplitude~\cite{Caneva2015Quantum,Shi2015multiphoton}. As a result, the dynamics of the atoms can be effectively described by a non-Hermitian Hamiltonian
\begin{equation}
\hat{H}_{{\rm eff}}=g_{{\rm eff}}\hat{\sigma}_{1}^{\dagger}\hat{\sigma}_{2}+g_{{\rm eff}}^{*}\hat{\sigma}_{2}^{\dagger}\hat{\sigma}_{1} -i\sum_{ij}\Gamma_{ij}\hat{\sigma}_{i}^{\dagger}\hat{\sigma}_{j}.\label{eq:Heff}
\end{equation}
with effective coupling strength in Eq.~(\ref{eq:geff}) and decay rates in Eq.~(\ref{eq:Gammaij}).
The time-evolution of an atomic operator $\hat{O}$ could be evaluated via
\begin{equation}
 e^{\mathcal{L}t}\hat{O}  =  e^{-i\hat{H}_{\rm eff} t}\hat{O}e^{i\hat{H}^{\dagger}_{\rm eff} t}. 
\end{equation}

Finally, we obtain the scattering amplitude
\begin{equation}
 \mathcal{A}_{r\rightarrow l}(-k,p) = \delta(k-p)r_{r\rightarrow l}(p)  
\end{equation}
where 
\begin{equation}
 r_{r\rightarrow l}(p) = \Gamma\sum_{ij}e^{ik_{0}(x_{i}+x_{j})}\left\langle gg\right|\hat{\sigma}_{i}\frac{i}{\hat{H}_{{\rm eff}}-p}\hat{\sigma}_{j}^{\dagger}\left|gg\right\rangle  
\end{equation}
is the reflection coefficient. In evaluating the atomic correlation function, we have utilized the fact 
$\langle gg| \exp\left(i\hat{H}^{\dagger}_{\rm eff}t\right) = \langle gg|$ and the relation
\begin{align}
& \int_{t_i}^{t_{2}}\!\!dt_{1}e^{-ip(t_{1}-t_{i})}{\rm Tr}_a\!\left[\hat{\sigma}_{i}e^{\mathcal{L}(t_{2}-t_{1})}\hat{\sigma}_{j}^{\dagger}|gg\rangle\langle gg|\right] \nonumber \\
= & \int_{t_i}^{t_{2}}\!\!dt_{1}e^{-ip(t_{1}-t_{i})}{\rm Tr}_a\left[\hat{\sigma}_{i}e^{-i\hat{H}_{\rm eff}(t_{2}-t_{1})}\hat{\sigma}_{j}^{\dagger}|gg\rangle\langle gg|e^{i\hat{H}^{\dagger}_{\rm eff}(t_2 - t_1)}\right] \nonumber\\
=&  \left\langle gg\right|\hat{\sigma}_{i}\frac{e^{-ip(t_2 - t_i)}}{i\hat{H}_{{\rm eff}}-ip}\hat{\sigma}_{j}^{\dagger}\left|gg\right\rangle\!, \!\! 
\end{align}
where term related to $\exp [-i\hat{H}_{\rm eff} (t_2 - t_i)]$ in the integral has been neglected, as it decays to zero for $t_i \rightarrow -\infty$. The corresponding wave function for $X<x_i$ of the output photon is given by
\begin{align}
 \Psi^{(1)}_{r\rightarrow l}(X,p,T) & = \langle gg|\otimes \langle 0|\hat{\psi}_l (X)e^{i\hat{H_0}t_f}e^{-i\hat{H}T} e^{-i\hat{H_0}t_i}\hat{b}_{p,r}^{\dagger}|0\rangle \otimes |gg\rangle\nonumber \\
 & = \frac{1}{\sqrt{2\pi}}e^{-i(k_{0}+p)X}r_{r\rightarrow l}(p). 
\end{align}
The output field is a plane wave photon with the same frequency as the input mode.

Similarly, we can obtain the scattering amplitude in the left-moving mode with  wavenumber $k_0+k$ as
\begin{align}
 \mathcal{A}_{r\rightarrow r}(k,p) & =\langle gg|\otimes \langle 0|\hat{b}_{k,r}e^{i\hat{H}_0 t_f}e^{-i\hat{H}T}e^{-i\hat{H}_0 t_i}\hat{b}_{p,r}|0\rangle\otimes |gg\rangle\nonumber\\
 &= \delta(k-p)t_{r\rightarrow r}(p)
\end{align}
with transmission coefficient
\begin{equation}
 t_{r\rightarrow r}(p) = 1+\Gamma\sum_{ij}e^{-ik_{0}(x_{i}-x_{j})}\left\langle gg\right|\hat{\sigma}_{i}\frac{i}{\hat{H}_{{\rm eff}}-p}\hat{\sigma}_{j}^{\dagger}\left|gg\right\rangle.    
\end{equation}
The wave function of the output plane single-photon for $X>x_i$ is given by
\begin{align}
 \Psi^{(1)}_{r\rightarrow r}(X,p,T) & \!=\! \langle gg|\!\otimes\! \langle 0|\hat{\psi}_r (X)e^{i\hat{H}_0 t_f}e^{-i\hat{H}T}e^{-i\hat{H}_0 t_i}\hat{b}_{p,r}^{\dagger}(t)|0\rangle \!\otimes \!|gg\rangle \nonumber\\
 & \!= \!\frac{1}{\sqrt{2\pi}}e^{-i(k_{0}+p)X}t_{r\rightarrow r}(p). 
\end{align}

\subsection{Scattering coefficients for a left-moving single photon input}
In this section, we consider the case where a left-moving plane-wave single photon with a wavenumber $-k_0-p$ is incident from the right side of the waveguide. The scattering amplitude in the right-moving mode with a wavenumber $k_0+k$ is given by:
\begin{align}
\mathcal{A}_{l\rightarrow r}(k,-p) & =  \langle gg|\otimes \langle 0|\hat{b}_{k,r}e^{i\hat{H}_0 t_f}e^{-i\hat{H}T}e^{-i\hat{H}_0 t_i}\hat{b}_{-p,l}^{\dagger}|0\rangle \otimes |gg\rangle \nonumber\\
& = \delta(k-p)r_{l\rightarrow r}(-p),
\end{align}
with reflection coefficient
\begin{equation}
 r_{l\rightarrow r}(-p) =  \Gamma \sum_{ij}e^{-ik_{0}(x_{i}+x_{j})}\left\langle gg\right|\hat{\sigma}_{i}\frac{i}{\hat{H}_{{\rm eff}}-p}\hat{\sigma}_{j}^{\dagger}\left|gg\right\rangle.   
\end{equation}
The corresponding wave function for $X>x_i$ reads
\begin{align}
 \Psi^{(1)}_{l\rightarrow r}(X,-p,T) & \!=\! \langle gg|\!\otimes\! \langle 0|\hat{\psi}_r (X)e^{i\hat{H}_0 t_f}e^{-i\hat{H}T}e^{-i\hat{H}_0 t_i}\hat{b}_{-p,l}^{\dagger}|0\rangle\! \otimes \!|gg\rangle\nonumber\\
 & = \frac{1}{\sqrt{2\pi}}e^{i(k_{0}+p)X}e^{-ipT}r_{l\rightarrow r}(-p).   
\end{align}

The scattering amplitude in the left-moving mode with a wavenumber $-k_0-k$ is given by:
\begin{align}
\mathcal{A}_{l\rightarrow l}(-k,-p) & =  \langle gg|\otimes \langle 0|\hat{b}_{-k,l}e^{i\hat{H}_0 t_f}e^{-i\hat{H}T}e^{-i\hat{H}_0 t_i}\hat{b}_{-p,l}^{\dagger}|0\rangle \otimes |gg\rangle \nonumber\\
& = \delta(k-p)t_{l\rightarrow l}(-p),
\end{align}
with transmission coefficient
\begin{equation}
 t_{l\rightarrow l}(-p) = 1 +\Gamma \sum_{ij}e^{ik_{0}(x_{i}-x_{j})}\left\langle gg\right|\hat{\sigma}_{i}\frac{i}{\hat{H}_{{\rm eff}}-p}\hat{\sigma}_{j}^{\dagger}\left|gg\right\rangle.   
\end{equation}
The corresponding wave function for $X<x_i$ reads
\begin{align}
 \Psi^{(1)}_{l\rightarrow l}(X,-p,T) & \!=\! \langle gg|\!\otimes\! \langle 0|\hat{\psi}_l (X)e^{i\hat{H}_0 t_f}e^{-i\hat{H}T}e^{-i\hat{H}_0 t_i} \hat{b}_{-p,l}^{\dagger}|0\rangle\! \otimes \!|gg\rangle\nonumber \\
 & = \frac{1}{\sqrt{2\pi}}e^{-i(k_{0}+p)X}t_{l\rightarrow l}(-p).   
\end{align}

\subsection{Output probabilities for two coherent inputs}
When the inputs of our beam splitter are weak coherent-state long pulses  ($|\alpha_{\lambda}|\ll1$), we can treat the pulses as quasi-single-mode pulses and expand the coherent states with low-excitation states [see Eq.~(\ref{eq:coherentstate})]
\begin{equation*}
\left|\alpha_{l}(-p)\right\rangle \!\!\approx\! e^{|\alpha_{l}|^{2}/2}\left[\!1\!+\!\!\sqrt{\epsilon}\alpha_{l} b_{-p,l}^{\dagger}\!+\!\frac{1}{2}(\!\sqrt{\epsilon}\alpha_{l})^{2}b_{-p,l}^{\dagger}b_{-p,l}^{\dagger} \!+ \!o(\alpha_l^3)\right]\left|0\right\rangle,
\end{equation*}
\begin{equation*}
\left|\alpha_{r}(p)\right\rangle \!\approx \!e^{|\alpha_{r}|^{2}/2}\left[\!1\!+\!\!\sqrt{\epsilon}\alpha_{r} b_{p,r}^{\dagger}\!+\!\frac{1}{2}(\!\sqrt{\epsilon}\alpha_{r})^{2}b_{p,r}^{\dagger}b_{p,r}^{\dagger}\! + \!o(\alpha_r^3)\right]\left|0\right\rangle\!.  
\end{equation*}
As shown in Sec.~\ref{sec2}, the two complex transmission coefficients have different arguments. Consequently, even when the amplitudes of the two coherent-state inputs are the same ($\alpha_r = \alpha_l=\alpha$), the output probabilities at the two ports are different in most cases as shown in the main text.

The mean photon number of the left and right output ports for $|X|>x_i$ up to the order of $|\alpha|^2$ are given by\begin{widetext}
\begin{align}
\left\langle \hat{\psi}_l^{\dagger}(X) \hat{\psi}_l(X) \right\rangle & \approx \epsilon|\alpha|^2\left|\Psi^{(1)}_{r\rightarrow l}(X,p,T)+\Psi^{(1)}_{l\rightarrow l}(X,-p,T)\right|^2 = \epsilon|\alpha|^2\left|r_{r\rightarrow l}(p) + t_{l\rightarrow l} (-p)\right|^2\\
\left\langle \hat{\psi}_r^{\dagger}(X) \hat{\psi}_r(X) \right\rangle & \approx \epsilon|\alpha|^2\left|\Psi^{(1)}_{l\rightarrow r}(X,-p,T)+\Psi^{(1)}_{r\rightarrow r}(X,p,T)\right|^2 = \epsilon|\alpha|^2\left|r_{l\rightarrow r}(-p) + t_{r\rightarrow r} (p)\right|^2 
\end{align}    
The normalized output probabilities are
\begin{align}
 P_l(p) & = \frac{\left\langle \hat{\psi}_l^{\dagger}(X) \hat{\psi}_l(X) \right\rangle}{2\epsilon|\alpha|^2}=\left|\frac{2i\Gamma (\Gamma\sin\theta+p\cos\theta+g\cos\phi)+g^{2}-p^{2}+2g\Gamma e^{i\phi}\sin\theta}{D(p)} \right|^2,\\   
 P_r(p) & = \frac{\left\langle \hat{\psi}_r^{\dagger}(X) \hat{\psi}_r(X) \right\rangle}{2\epsilon|\alpha|^2}=\left|\frac{2i\Gamma(\Gamma\sin\theta+p\cos\theta+g\cos\phi)+g^{2}-p^{2}+2g\Gamma e^{-i\phi}\sin\theta}{D(p)} \right|^2. 
\end{align}   
\end{widetext}

\subsection{Scattering channels analysis \label{sec:scatteringchannels}}
In the previous sections, we demonstrated that the reflection and transmission coefficients are primarily determined by the effective Hamiltonian $\hat{H}_{\rm eff}$ of the atoms, as given in Eq.~(\ref{eq:Heff}). Here, we provide a detailed analysis of $\hat{H}_{\rm eff}$ and the scattering channels. To begin, we introduce two new single-excitation states.
\begin{equation}
\left|\pm\right\rangle =\frac{1}{\sqrt{2}}\left(\left|eg\right\rangle \pm\left|ge\right\rangle \right)
\end{equation}
with positive and negative parity, respectively. We can also define the corresponding ladder operators 
\begin{equation}
\hat{\sigma}_{\pm}=\frac{1}{\sqrt{2}}\left(\hat{\sigma}_{1}\pm\hat{\sigma}_{2}\right),\ \hat{\sigma}_{\pm}^{\dagger}=\frac{1}{\sqrt{2}}\left(\hat{\sigma}_{1}^{\dagger}\pm\hat{\sigma}_{2}^{\dagger}\right).
\end{equation}
Their actions on the double-excitation state are slightly different from $\hat{\sigma}_i$ in the original representation
\begin{align}
\left(\hat{\sigma}_{+}^{\dagger}\right)^{2}\left|gg\right\rangle  & =\left|ee\right\rangle ,\ \left(\hat{\sigma}_{-}^{\dagger}\right)^{2}\left|gg\right\rangle =-\left|ee\right\rangle \\
\hat{\sigma}_{+}^{\dagger}\hat{\sigma}_{-}^{\dagger}\left|gg\right\rangle  & =\hat{\sigma}_{-}^{\dagger}\hat{\sigma}_{+}^{\dagger}\left|gg\right\rangle =0\\
\hat{\sigma}_{+}\left|ee\right\rangle  & =\left|+\right\rangle ,\ \hat{\sigma}_{-}\left|ee\right\rangle =-\left|-\right\rangle .
\end{align}

The matrix elements of the effective Hamiltonian in the basis $\{\left|ee\right\rangle ,\left|+\right\rangle ,\left|-\right\rangle ,\left|gg\right\rangle \}$ are given by
\begin{equation}
H_{{\rm eff}}=\left[\begin{array}{cccc}
-2i\Gamma & 0 & 0 & 0\\
0 & -i\alpha_{+} & -i\beta & 0\\
0 & i\beta & -i\alpha_{-} & 0\\
0 & 0 & 0 & 0
\end{array}\right],\label{eq:Heff_mat}
\end{equation}
with
\begin{align}
\alpha_{+} & =\Gamma(1+e^{i\theta})+ig\cos\phi\\
\alpha_{-} & =\Gamma(1-e^{i\theta})-ig\cos\phi\\
\beta & =g\sin\phi.
\end{align}
The Hamiltonian $\hat{H}_{\rm eff}$ has not been diagonalized with the $|\pm\rangle$ states. However, the dissipation terms only exist in the diagonal elements. Thus, states $|\pm\rangle$ determine the decay channels and the off-diagonal terms induce coherent interaction between these channels.

In deriving the analytical expression of the scattering coefficients and multi-time correlation function of photons, we also need the following three operators:
\begin{align}
\hat{M} & \equiv \frac{-i}{\hat{H}_{\rm eff}-p}, \label{eq:M} \\   
\hat{U}(\tau) & \equiv \exp (-i\hat{H}_{\rm eff}\tau),\label{eq:U}\\
\hat{N}(\tau) & \equiv  \exp (-i\hat{H}_{\rm eff}\tau) \frac{-i}{\hat{H}_{\rm eff}-p}.
\end{align}
Their matrix elements in the basis $\{\left|ee\right\rangle ,\left|+\right\rangle ,\left|-\right\rangle ,\left|gg\right\rangle \}$ are given by
\begin{equation}
M=\left[\begin{array}{cccc}
\frac{1}{2\Gamma-ip} & 0 & 0 & 0\\
0 & \frac{\alpha_{-}-ip}{D(p)} & \frac{-\beta}{D(p)} & 0\\
0 & \frac{\beta}{D(p)} & \frac{\alpha_{+}-ip}{D(p)} & 0\\
0 & 0 & 0 & \frac{i}{p}
\end{array}\right],\label{eq:M_mat}
\end{equation}
\begin{equation}
U(\tau)=\left[\begin{array}{cccc}
e^{-2\Gamma\tau} & 0 & 0 & 0\\
0 & U_{22}(\tau) & U_{23}(\tau) & 0\\
0 & U_{32}(\tau) & U_{33}(\tau) & 0\\
0 & 0 & 0 & 1
\end{array}\right],
\end{equation}
\begin{equation}
 N(\tau)=\left[\begin{array}{cccc}
\frac{1}{\Gamma-ip}e^{-2\Gamma\tau} & 0 & 0 & 0\\
0 & N_{22}(\tau) & N_{23}(\tau) & 0\\
0 & N_{32}(\tau) & N_{33}(\tau) & 0\\
0 & 0 & 0 & \frac{i}{p}
\end{array}\right],   
\end{equation}
where
\begin{equation}
 D(p) = (\alpha_{+}-ip)(\alpha_{-}-ip)+\beta^{2},\label{eq:D}   
\end{equation}
is the determinant of the matrix $iH-ip$. The rest elements of $U$ and $N$ are given by 
\begin{align}
U_{22}(\tau) & =\left[\frac{\alpha_{-}-\alpha_{+}}{\sqrt{\Delta}}\sin\left(\frac{\sqrt{\Delta}}{2}\tau\right)+\cos\left(\frac{\sqrt{\Delta}}{2}\tau\right)\right]e^{-\Gamma\tau},\\
U_{33}(\tau) & =\left[-\frac{\alpha_{-}-\alpha_{+}}{\sqrt{\Delta}}\sin\left(\frac{\sqrt{\Delta}}{2}\tau\right)+\cos\left(\frac{\sqrt{\Delta}}{2}\tau\right)\right]e^{-\Gamma\tau},\\
U_{23}(\tau) & =-U_{32}(\tau)=-\frac{2\beta}{\sqrt{\Delta}}\sin\left(\frac{\sqrt{\Delta}}{2}\tau\right)e^{-\Gamma\tau},
\end{align}
\begin{widetext}
\begin{align}
N_{22}(\tau) & = \frac{(\alpha_{-}-ip)\sqrt{\Delta}\cos\left(\frac{\sqrt{\Delta}}{2}\tau\right)-[(\alpha_{+}-\alpha_{-})(\alpha_{-}-ip)+2\beta^{2}]\sin\left(\frac{\sqrt{\Delta}}{2}\tau\right)}{\sqrt{\Delta}D(p)}e^{-\Gamma\tau},\\
N_{33}(\tau) & = \frac{(\alpha_{+}-ip)\sqrt{\Delta}\cos\left(\frac{\sqrt{\Delta}}{2}\tau\right)-[(\alpha_{+}-\alpha_{-})(\alpha_{+}-ip)+2\beta^{2}]\sin\left(\frac{\sqrt{\Delta}}{2}\tau\right)}{\sqrt{\Delta}D(p)}e^{-\Gamma\tau},\\
N_{23} & = -N_{32}=-\frac{\beta\left[\sqrt{\Delta}\cos\left(\frac{\sqrt{\Delta}\tau}{2}\right)+(\alpha_{+}+\alpha_{-}-2ip)\sin\left(\frac{\sqrt{\Delta}\tau}{2}\right)\right]}{\sqrt{\Delta}D(p)}e^{-\Gamma\tau},
\end{align}\end{widetext}
where 
\begin{equation}
 \Delta = -(\alpha_{-}-\alpha_{+})^{2}+4\beta^{2} = 4\left(g\cos\phi-i\Gamma e^{i\theta}\right)^{2}+4g\sin^{2}\phi,  
\end{equation}
is the discriminant of the quadratic eigenvalue equation of the matrix $H_{\rm eff}$.

All four matrices $H_{\rm eff}$, $M$, $U$, and $N$ share the property that their off-diagonal elements are anti-symmetric, specifically, $O_{23}=-O_{32}$. As demonstrated later, this results in perfect destructive (constructive) interference between the scattering channels for the reflection (transmission) coefficient. It is important to note that only the off-diagonal elements of these four matrices contain the parity symmetry-breaking term, i.e., the odd powers of $\beta = g\sin\phi$.

\section{Two-photon scattering processes for two coherent inputs\label{Appendix4}}
In this section, we study the statistical properties of the output photons. We mainly focus on the $g^{(2)}$-function, which defined as
\begin{align}
g_{\lambda}^{(2)}(\tau) & =\frac{\left\langle \hat{\psi}_{\lambda}^{\dagger}(X)\hat{\psi}_{\lambda}^{\dagger}(X+\lambda\tau)\hat{\psi}_{\lambda}(X+\lambda\tau)\hat{\psi}_{\lambda}(X)\right\rangle }{\left\langle \hat{\psi}_{\lambda}^{\dagger}(X)\hat{\psi}_{\lambda}(X)\right\rangle \left\langle \hat{\psi}_{\lambda}^{\dagger}(X+\lambda\tau)\hat{\psi}_{\lambda}(X+\lambda\tau)\right\rangle}.
\end{align}
For two weak coherent-state inputs ($|\alpha_l(-p)\rangle\otimes|\alpha_{r}(p)\rangle$), only the three two-photon components
\begin{equation}
\propto\left[b_{p,r}^{\dagger}b_{-p,l}^{\dagger}+\frac{1}{2}b_{-p,l}^{\dagger}b_{-p,l}^{\dagger}+\frac{1}{2}b_{p,r}^{\dagger}b_{p,r}^{\dagger}\right]\left|0\right\rangle    
\end{equation} 
will contribute to the second-order coherence function. Next, we use the scattering method to evaluate the two-photon wave functions corresponding to the three input states, separately.

\subsection{Two photons incident from the left \label{sec: twophotonleft}}
In this section, we examine the scattering of two right-moving photons incident from the left side of the beam splitter. There are three possible output states: (1) two reflected photons coming out from the left side of the waveguide; (2) two transmitted photons coming out from the right side of the waveguide; (3) one reflected and one transmitted photon coming out from both ends of the waveguide. The third case will not contribute to the $g^{(2)}$-function of the output, and we will neglect it in the following discussion.

\subsubsection{Wave function of two left-moving photons}
The wave function of the left-moving two photons with $X<x_{i}$ and $\tau\geq 0$ in the limit $T=t_f -t_i\rightarrow\infty$ is obtained
\begin{widetext}
\begin{align}
\Psi^{(2)}_{r\rightarrow l}(X-\tau,X) = & \left\langle gg,0\right|\hat{\psi}_{l}(X)\hat{\psi}_{l}(X-\tau)e^{i\hat{H}_0 t_f}e^{-i\hat{H}T}e^{-i\hat{H}_0 t_i}\hat{b}_{p,r}^{\dagger}\hat{b}_{p,r}^{\dagger}\left|gg,0\right\rangle \\
=  & \frac{1}{2\pi}e^{-i(k_{0}+p)(2X-\tau)}\Gamma^{2}\sum_{ijmn}2e^{ik_{0}(x_{i}+x_{j}+x_{m}+x_{n})} \left[\left\langle gg\right|\hat{\sigma}_{i}e^{-i\hat{H}_{{\rm eff}}\tau}\hat{\sigma}_{j}\frac{1}{i\hat{H}_{{\rm eff}}-2ip}\hat{\sigma}_{m}^{\dagger}\frac{1}{i\hat{H}_{{\rm eff}}-ip}\hat{\sigma}_{n}^{\dagger}\left|gg\right\rangle \right.\nonumber \\
& \left.+\left\langle gg\right|\hat{\sigma}_{i}\left(e^{-ip\tau}-e^{-i\hat{H}_{{\rm eff}}\tau}\right)\frac{1}{i\hat{H}_{{\rm eff}}-ip}\hat{\sigma}_{m}^{\dagger}\left|gg\right\rangle \left\langle gg\right|\hat{\sigma}_{j}\frac{1}{i\hat{H}_{{\rm eff}}-ip}\hat{\sigma}_{n}^{\dagger}\left|gg\right\rangle \right], \label{eq:psirl1}
\end{align}    
\end{widetext}
where we have used the input-output relation and the quantum regression theorem. In the following, we first consider two special cases with $\tau=0$ and $\tau\rightarrow\infty$, respectively. The two-photon wave function is exchange symmetric $\Psi^{(2)}_{r\rightarrow l}(X-\tau,X)=\Psi^{(2)}_{r\rightarrow l}(X,X-\tau)$, since two field operators $\hat{\psi}_l (X)$ and $\hat{\psi}_l (X-\tau)$ commute. 

For $\tau=0$, the second term in Eq.~(\ref{eq:psirl1}) vanishes and the two-photon wave function reads
\begin{equation}
\left.\Psi^{(2)}_{r\rightarrow l} (X-\tau,X)\right|_{\tau =0}=\frac{1}{2\pi}e^{-2i(k_{0}+p)X}\frac{2\Gamma}{\Gamma-ip}[1-t_{r\rightarrow r}(p)], 
\end{equation}
where we have used the facts\begin{widetext}
\begin{align}
\left\langle gg\right|\hat{\sigma}_{i}e^{-i\hat{H}_{{\rm eff}}\tau}\hat{\sigma}_{j}\frac{1}{i\hat{H}_{{\rm eff}}-2ip}\hat{\sigma}_{m}^{\dagger}\frac{1}{i\hat{H}_{{\rm eff}}-ip}\hat{\sigma}_{n}^{\dagger}\left|gg\right\rangle =\left\langle gg\right|\hat{\sigma}_{i}e^{-i\hat{H}_{{\rm eff}}\tau}\hat{\sigma}_{j}\frac{1}{i\hat{H}_{{\rm eff}}-2ip}|ee\rangle\langle ee|\hat{\sigma}_{m}^{\dagger}\frac{1}{i\hat{H}_{{\rm eff}}-ip}\hat{\sigma}_{n}^{\dagger}\left|gg\right\rangle,     
\end{align}    
\begin{align}
\Gamma\sum_{mn}e^{ik_0(x_m+x_n)}\langle ee|\hat{\sigma}_{m}^{\dagger}\frac{1}{i\hat{H}_{{\rm eff}}-ip}\hat{\sigma}_{n}^{\dagger}\left|gg\right\rangle & = 2\Gamma\left\langle ee\right|\left(\hat{\sigma}_{+}^{\dagger}\cos\frac{\theta}{2}-i\hat{\sigma}_{-}\sin\frac{\theta}{2}\right)M\left(\hat{\sigma}_{+}^{\dagger}\cos\frac{\theta}{2}-i\hat{\sigma}_{-}^{\dagger}\sin\frac{\theta}{2}\right)\left|gg\right\rangle \\
& = \Gamma\left[2\left(M_{22}\cos^{2}\frac{\theta}{2}+M_{33}\sin^{2}\frac{\theta}{2}\right)-i(M_{23}-M_{32})\sin\theta\right]=  \left[1-t_{r\rightarrow r}(p)\right].
\end{align}
We see that perfect constructive interference (the $M_{23}-M_{32}$ term) between the double-excitation channels $|gg\rangle\rightarrow |+\rangle\rightarrow |-\rangle\rightarrow |ee\rangle$ and $|gg\rangle\rightarrow |-\rangle\rightarrow |+\rangle\rightarrow |ee\rangle$ occurs. For the case $\tau\rightarrow\infty$, the terms containing $\exp (-i\hat{H}_{\rm eff} \tau)$ decay to zero exponentially. Only the term which describes the two photons are reflected separately by the atoms, contributes to the wave function
\begin{equation}
\left.\Psi^{(2)}_{r\rightarrow l} (X-\tau,X)\right|_{\tau\rightarrow\infty} =\frac{1}{2\pi}e^{-i(k_{0}+p)(2X-\tau)-2ipT}\times 2r_{r\rightarrow l}^{2}(p). 
\end{equation}

The wave function of the two reflected photons for an arbitrary $\tau$ is given by
\begin{align}
\Psi^{(2)}_{r\rightarrow l}(X-\tau,X)= & \frac{1}{2\pi}e^{-i(k_{0}+p)(2X-\tau)}\left\{ 2e^{-ip\tau}r_{r\rightarrow l}^{2}(p) +\frac{\Gamma}{\Gamma-ip}\left[2\left(U_{22}\cos^{2}\frac{\theta}{2}+U_{33}\sin^{2}\frac{\theta}{2}\right)+i(U_{23}-U_{32})\sin\theta\right][1-t_{r\rightarrow r}(p)]\nonumber \right.\\
 & \left.+2\Gamma\left[2\left(N_{22}\cos^{2}\frac{\theta}{2}-N_{33}\sin^{2}\frac{\theta}{2}\right)-i(N_{23}+N_{32})\sin\theta\right]r_{r\rightarrow l}(p)\right\} \\
 = & \frac{1}{2\pi}e^{-i(k_{0}+p)(2X-\tau)}\left\{ \frac{\Gamma}{\Gamma-ip}\left[2\cos\left(\frac{\sqrt{\Delta}}{2}\tau\right)+\frac{2(\alpha_{-}-\alpha_{+})\cos\theta-4i\beta\sin\theta}{\sqrt{\Delta}}\sin\left(\frac{\sqrt{\Delta}}{2}\tau\right)\right][1-t_{r\rightarrow r}(p)]e^{-\Gamma\tau}\right.\nonumber \\
 & \left.-2\left[r_{r\rightarrow l}(p)\cos\left(\frac{\sqrt{\Delta}\tau}{2}\right)-2\Gamma\frac{(\alpha_{-}-\alpha_{+})(\Gamma-ip)-\Delta\cos\theta}{\sqrt{\Delta}D(p)}\sin\left(\frac{\sqrt{\Delta}\tau}{2}\right)\right]r_{r\rightarrow l}(p)e^{-\Gamma\tau}+2e^{-ip\tau}r_{r\rightarrow l}^{2}(p)\right\}. 
\end{align}    
Here, we see that perfect constructive interference (the $U_{23}-U_{32}$ term) between the scattering channels occurs between the de-excitation channels $|ee\rangle\rightarrow |+\rangle\rightarrow |-\rangle\rightarrow |gg\rangle$ and $|ee\rangle\rightarrow |-\rangle\rightarrow |+\rangle\rightarrow |gg\rangle$ and perfect destructive interference (the $N_{23}+N_{32}$ term) occurs between the single-photon scattering channels $|gg\rangle\rightarrow |+\rangle\rightarrow |-\rangle\rightarrow |gg\rangle$ and $|gg\rangle\rightarrow |-\rangle\rightarrow |+\rangle\rightarrow |gg\rangle$.
The existing perfect constructive interference will break the parity of both the $g^{(2)}(0)$- and $g^{(2)}(\tau)$-functions.

\subsubsection{Wave function of two right-moving photons}
The wave function of the right-moving two photons with $x_{i}<X$ and $\tau \geq 0$
in the limit $T\rightarrow\infty$ is given by
\begin{align}
\Psi^{(2)}_{r\rightarrow r}(X+\tau,X)= & \left\langle gg,0\right|\hat{\psi}_{r}(X)\hat{\psi}_{r}(X+\tau)e^{i\hat{H}_0 t_f}e^{-i\hat{H}T}e^{-i\hat{H}_0 t_i}\hat{b}_{p,r}^{\dagger}(t_{i})\hat{b}_{p,r}^{\dagger}(t_{i})\left|gg,0\right\rangle \\
= & \frac{1}{2\pi}e^{i(k_{0}+p)(2X+\tau)}\left\{ 2e^{-ip\tau} -4\Gamma\sum_{ij}e^{-ik_{0}(x_{i}-x_{j})-ip\tau}\left\langle gg\right|\hat{\sigma}_{i}\frac{1}{i\hat{H}_{{\rm eff}}-ip}\hat{\sigma}_{j}^{\dagger}\left|gg\right\rangle \right.\nonumber \\
& +2\Gamma^{2}\sum_{ijmn}e^{-ik_{0}(x_{i}+x_{j}-x_{m}-x_{n})}\left[\left\langle gg\right|\hat{\sigma}_{i}e^{-i\hat{H}_{{\rm eff}}\tau}\hat{\sigma}_{j}\frac{1}{i\hat{H}_{{\rm eff}}-2ip}\hat{\sigma}_{m}^{\dagger}\frac{1}{i\hat{H}_{{\rm eff}}-ip}\hat{\sigma}_{n}^{\dagger}\left|gg\right\rangle\right.\nonumber\\
& \left.\left.+\left\langle gg\right|\hat{\sigma}_{i}\left(e^{-ip\tau}-e^{-i\hat{H}_{{\rm eff}}\tau}\right)\frac{1}{i\hat{H}_{{\rm eff}}-ip}\hat{\sigma}_{n}^{\dagger}\left|gg\right\rangle \left\langle gg\right|\hat{\sigma}_{j}\frac{1}{i\hat{H}_{{\rm eff}}-ip}\hat{\sigma}_{m}^{\dagger}\left|gg\right\rangle \right]\right\}.
\end{align}    
For the case $\tau = 0$, we have
\begin{equation}
 \left.\Psi^{(2)}_{r\rightarrow r}(X+\tau,X)\right|_{\tau=0}= \frac{1}{2\pi}e^{2i(k_{0}+p)X}\times 2\left[2t_{r\rightarrow r}(p)-1+\frac{\Gamma}{\Gamma-ip}[1-t_{r\rightarrow r}(p)]\right]   
\end{equation}
For the case $\tau\rightarrow\infty$, we have
\begin{equation}
 \left. \Psi^{(2)}_{r\rightarrow r}(X+\tau,X)\right|_{\tau\rightarrow\infty} =\frac{1}{2\pi}e^{i(k_{0}+p)(2X+\tau)}\times2t_{r\rightarrow r}^{2}(p),   
\end{equation}
describing the fact that two photons are transmitted independently.

For an arbitrary $\tau$, we have
\begin{align}
\Psi^{(2)}_{r\rightarrow r}(X+\tau,X) = &  \frac{1}{2\pi}e^{i(k_{0}+p)(2X+\tau)}\left\{ 2e^{-ip\tau}t_{r\rightarrow r}^{2}(p)+\frac{\Gamma[1-t_{r\rightarrow r}(p)]}{\Gamma-ip}\left[2\left(U_{22}\cos^{2}\frac{\theta}{2}+U_{33}\sin^{2}\frac{\theta}{2}\right)-i(U_{23}-U_{32})\sin\theta\right]\right.\nonumber \\
 & \left.-2\Gamma\left[2\left(N_{22}\cos^{2}\frac{\theta}{2}+N_{33}\sin^{2}\frac{\theta}{2}\right)-i(N_{23}-N_{32})\sin\theta\right][1-t_{r\rightarrow r}(p)]\right\} \\
= & \frac{1}{2\pi}e^{i(k_{0}+p)(2X+\tau)}\left\{ 2e^{-ip\tau}t_{r\rightarrow r}^{2}(p)+\frac{\Gamma [1-t_{r\rightarrow r}(p)]}{\Gamma-ip}\left[2\cos\left(\frac{\sqrt{\Delta}}{2}\tau\right)+\frac{2(\alpha_{-}-\alpha_{+})\cos\theta+4i\beta\sin\theta}{\sqrt{\Delta}}\sin\left(\frac{\sqrt{\Delta}}{2}\tau\right)\right]e^{-\Gamma\tau}\nonumber \right.\\
 & \left.-2\left[[1-t_{r\rightarrow r}(p)]\cos\left(\frac{\sqrt{\Delta}\tau}{2}\right)+2\Gamma\frac{[(\alpha_{-}-\alpha_{+})\cos\theta+2i\beta\sin\theta](\Gamma-ip)-\Delta}{\sqrt{\Delta}D(p)}\right][1-t_{r\rightarrow r}(p)]e^{-\Gamma\tau}\right\}. 
\end{align}
Here we see that two perfect constructive interferences $U_{23}-U_{32}$ and $N_{23}-N_{32}$ occur.

\subsection{Two photons incident from the right}
In this section, we check the scattering of two left-moving photons incident from the right end of the waveguide. The scattering processes are similar to Sec~\ref{sec: twophotonleft}. We only list the final results.

\subsubsection{Wave function of two left-moving photons}
The wave function of the right-moving two photons with $X<x_{i}$ and $\tau\geq 0$ is given by
\begin{align}
\Psi^{(2)}_{l\rightarrow l}&(X-\tau,X) =  \left\langle gg,0\right|\hat{\psi}_{r}(X)\hat{\psi}_{r}(X-\tau)e^{i\hat{H}_0 t_f}e^{-i\hat{H}T}e^{-i\hat{H}_0 t_i}\hat{b}_{-p,l}^{\dagger}\hat{b}_{-p,l}^{\dagger}\left|gg,0\right\rangle \\
= &  \frac{1}{2\pi}e^{-i(k_{0}+p)(2X-\tau)-2ipT}\left\{ 2e^{-ip\tau}t_{l\rightarrow l}^{2}(-p)+\frac{\Gamma[1\!-\!t_{l\rightarrow l}(-p)]}{\Gamma-ip}\left[2\cos\left(\frac{\sqrt{\Delta}}{2}\tau\right)\!+\!\frac{2(\alpha_{-}\!-\!\alpha_{+})\cos\phi-4i\beta\sin\theta}{\sqrt{\Delta}}\sin\left(\frac{\sqrt{\Delta}}{2}\tau\right)\right]e^{-\Gamma\tau}\right.\nonumber \\
 & \left.-2\left[[1\!-\!t_{l\rightarrow l}(-p)]\cos\left(\!\frac{\sqrt{\Delta}\tau}{2}\!\right)\!+\!\Gamma\frac{[(\alpha_{-}\!-\!\alpha_{+})\cos\theta-2i\beta\sin\theta](\alpha_{+}\!+\!\alpha_{-}\!-\!2ip)\!-\!\Delta}{\sqrt{\Delta}D(p)}\sin\left(\!\frac{\sqrt{\Delta}\tau}{2}\!\right)\right][1\!-\!t_{l\rightarrow l}(-p)]e^{-\Gamma\tau}\!\right\}\!. \!
\end{align}
For the case $\tau = 0$
\begin{equation}
 \left.\Psi^{(2)}_{l\rightarrow l}(X-\tau,X)\right|_{\tau=0} = \frac{1}{2\pi}e^{-2i(k_{0}+p)X}\times2\left[2t_{l\rightarrow l}(-p)-1+\frac{\Gamma}{\Gamma-ip}[1-t_{l\rightarrow l}(-p)]\right].  
\end{equation}
For the case $\tau\rightarrow\infty$, we have
\begin{equation}
\left.\Psi^{(2)}_{l\rightarrow l}(X-\tau,X)\right|_{\tau\rightarrow\infty} =\frac{1}{2\pi}e^{-i(k_{0}+p)(2X-\tau)}\times 2 t_{l\rightarrow l}^{2}(p).
\end{equation}

\subsubsection{Wave function of two right-moving photons}
The wave function of the right-moving two photons with $X>x_{i}$ and $\tau\geq 0$ is given by
\begin{align}
\Psi^{(2)}_{l\rightarrow r}&(X+\tau,X) =  \left\langle gg,0\right|\hat{\psi}_{l}(X)\hat{\psi}_{l}(X+\tau)e^{i\hat{H}_0 t_f}e^{-i\hat{H}T}e^{-i\hat{H}_0 t_i}\hat{b}_{-p,l}^{\dagger}\hat{b}_{-p,l}^{\dagger}\left|gg,0\right\rangle \\
= & \frac{1}{2\pi}e^{i(k_{0}+p)(2X+\tau)}\left\{ 2e^{-ip\tau}r_{l\rightarrow r}^{2}(-p) +\frac{\Gamma[1-t_{l\rightarrow l}(-p)]}{\Gamma-ip}\left[2\cos\left(\frac{\sqrt{\Delta}}{2}\tau\right)+\frac{2(\alpha_{-}-\alpha_{+})\cos\theta+4i\beta\sin\theta}{\sqrt{\Delta}}\sin\left(\frac{\sqrt{\Delta}}{2}\tau\right)\right]e^{-\Gamma\tau}\right. \nonumber \\
 & \left.-2\left[r_{l\rightarrow r}(-p)\cos\left(\frac{\sqrt{\Delta}\tau}{2}\right)-2\Gamma\frac{(\alpha_{-}-\alpha_{+})(\Gamma-ip)-\Delta\cos\theta}{\sqrt{\Delta}D(p)}\sin\left(\frac{\sqrt{\Delta}\tau}{2}\right)\right]r_{l\rightarrow r}(-p)e^{-\Gamma\tau}\right\} .
\end{align}
For the case $\tau =0$, we have
\begin{equation}
 \left.\Psi^{(2)}_{l\rightarrow r}(X+\tau,X)\right|_{\tau=0} = \frac{1}{2\pi}e^{2i(k_{0}+p)X}\frac{2\Gamma}{\Gamma-ip}[1-t_{l\rightarrow l}(-p)].  
\end{equation}
For the case $\tau\rightarrow\infty$, we have
\begin{equation}
\left.\Psi^{(2)}_{l\rightarrow r}(X+\tau,X)\right|_{\tau\rightarrow\infty} =\frac{1}{2\pi}e^{i(k_{0}+p)(2X+\tau)}\times 2 r_{r\rightarrow r}^{2}(-p).
\end{equation}

\subsection{Two photons incident from two ends \label{sec:TwoSides}}
When two photons are incident from the two ends of the waveguide, we also only consider the cases with two photons coming out from the same output port.

\subsubsection{Wave function of two left-moving photons}
The wave function of the left-moving two photons with $X<x_{i}$ and $\tau\geq 0$ is obtained
\begin{align}
 \Psi^{(2)}_{{\rm TS}\rightarrow l}(X-\tau,X) = & \left\langle gg,0\right|\hat{\psi}_{l}(X)\hat{\psi}_{l}(X-\tau)e^{i\hat{H}_0 t_f}e^{-i\hat{H}T}e^{-i\hat{H}_0 t_i}\hat{b}_{-p,l}^{\dagger}\hat{b}_{p,r}^{\dagger}\left|gg,0\right\rangle \\
= & \frac{1}{2\pi}e^{-i(k_{0}+p)(2X-\tau)}\left\{ 2e^{-ip\tau}r_{r\rightarrow l}(p)t_{l\rightarrow l}(-p)+2[1-t_{l\rightarrow l}(-p)]r_{r\rightarrow l}(p)\cos\left(\frac{\sqrt{\Delta}\tau}{2}\right)e^{-\Gamma\tau}\right.\nonumber \\
 & -\frac{\Gamma}{\Gamma-ip}\left[2\cos\left(\frac{\sqrt{\Delta}}{2}\tau\right)+\frac{2(\alpha_{-}-\alpha_{+})\cos\theta-4i\beta\sin\theta}{\sqrt{\Delta}}\sin\left(\frac{\sqrt{\Delta}}{2}\tau\right)\right]r_{r\rightarrow l}(p)e^{-\Gamma\tau}\nonumber \\
 & +2\Gamma\frac{[(\alpha_{-}-\alpha_{+})\cos\theta-2i\beta\sin\theta](\Gamma -ip)-\Delta}{\sqrt{\Delta}D(p)}\sin\left(\frac{\sqrt{\Delta}\tau}{2}\right)r_{r\rightarrow l}(p)e^{-\Gamma\tau}\nonumber \\
 & -2\Gamma\frac{(\alpha_{-}-\alpha_{+})(\Gamma-ip)-\Delta\cos\theta}{\sqrt{\Delta}D(p)}\sin\left(\frac{\sqrt{\Delta}\tau}{2}\right)[1-t_{l\rightarrow l}(-p)]e^{-\Gamma\tau}
\end{align}
For the case $\tau = 0$, we have
\begin{equation}
 \left.\Psi^{(2)}_{{\rm TS}\rightarrow l} \right|_{\tau = 0}=\frac{1}{2\pi}e^{-2i(k_{0}+p)X}\left[2-\frac{2\Gamma}{(\Gamma-ip)}\right]r_{r\rightarrow l}(p).    
\end{equation}
For the case $\tau\rightarrow\infty$, we have
\begin{equation}
\left.\Psi^{(2)}_{{\rm TS}\rightarrow l} \right|_{\tau\rightarrow\infty}=\frac{1}{2\pi}e^{-i(k_{0}+p)(2X-\tau)}\times 2 r_{r\rightarrow l}(p)t_{l\rightarrow l}(-p).     
\end{equation}

\subsubsection{Wave function of two right-moving photons}
The wave function of the right-moving two photons with $X<x_{i}$ and $\tau\geq 0$ is given by
\begin{align}
\Psi^{(2)}_{{\rm TS}\rightarrow r}(X+\tau,X) = & \left\langle G,0\right|\hat{\psi}_{r}(X)\hat{\psi}_{r}(X+\tau)e^{i\hat{H}_0 t_f}e^{-i\hat{H}T}e^{-i\hat{H}_0 t_i}\hat{b}_{-p,l}^{\dagger}\hat{b}_{p,r}^{\dagger}\left|G,0\right\rangle\\ 
= & \frac{1}{2\pi}e^{i(k_{0}+p)(2X+\tau)}\left\{ 2e^{-ip\tau}r_{l\rightarrow r}(-p)t_{r\rightarrow r}(p)+2r_{l\rightarrow r}(-p)[1-t_{r\rightarrow r}(p)]\cos\left(\frac{\sqrt{\Delta}\tau}{2}\right)e^{-\Gamma\tau}\right.\nonumber \\
 & -\frac{\Gamma}{\Gamma-ip}\left[2\cos\left(\frac{\sqrt{\Delta}}{2}\tau\right)+\frac{2(\alpha_{-}-\alpha_{+})\cos\theta+4i\beta\sin\theta}{\sqrt{\Delta}}\sin\left(\frac{\sqrt{\Delta}}{2}\tau\right)\right]r_{l\rightarrow r}(-p)e^{-\Gamma\tau}\nonumber \\
 & +2\Gamma\frac{[(\alpha_{-}-\alpha_{+})\cos\theta+2i\beta\sin\theta](\Gamma-ip)-\Delta}{\sqrt{\Delta}D(p)}\sin\left(\frac{\sqrt{\Delta}\tau}{2}\right)r_{l\rightarrow r}(-p)e^{-\Gamma\tau}\nonumber \\
 & \left.-2\Gamma\frac{(\alpha_{-}-\alpha_{+})(\Gamma - ip)-\Delta\cos\theta}{\sqrt{\Delta}D(p)}\sin\left(\frac{\sqrt{\Delta}\tau}{2}\right)[1-t_{r\rightarrow r}(p)]e^{-\Gamma\tau}\right\}.
\end{align}
For the case $\tau = 0$, we have
\begin{equation}
 \left.\Psi^{(2)}_{{\rm TS}\rightarrow r} (X+\tau,X)\right|_{\tau =0} = \frac{1}{2\pi}e^{2i(k_{0}+p)X}\left[2-\frac{2\Gamma}{(\Gamma-ip)}\right]r_{l\rightarrow r}(-p).   
\end{equation}
For the case $\tau\rightarrow\infty$, we have
\begin{equation}
\left.\Psi^{(2)}_{{\rm TS}\rightarrow r} (X+\tau,X)\right|_{\tau =0} = \frac{1}{2\pi}e^{i(k_{0}+p)(2X+\tau)}\times 2 r_{l\rightarrow r}(-p)t_{r\rightarrow r}(p).    
\end{equation}

\subsection{Second-order coherence function}
When two weak coherent-state drivings are applied to the two ends of our beam splitter, only the two-photon components contribute to the $g^{(2)}$-function. Consequently, the analytical results of the coherence functions can be obtained with the output two-photon wave functions
\begin{equation}
 g^{(2)}_{l} (\tau) = \frac{(2\pi)^2|\Psi^{(2)}_{l}(X-\tau,X)|^{2}}{\left|r_{r\rightarrow l}(p)+t_{l \rightarrow l}(-p)\right|^{4}}, \ g^{(2)}_{r} (\tau) = \frac{(2\pi)^2|\Psi^{(2)}_{r}(X+\tau,X)|^{2}}{\left|r_{l\rightarrow r}(-p)+t_{r \rightarrow }(p)\right|^{4}},  
\end{equation}
where
\begin{align}
 \Psi_{l}^{(2)}&(X-\tau,X) = \Psi^{(2)}_{{\rm TS}\rightarrow l}(X-\tau,X)+\frac{1}{2}\Psi^{(2)}_{r\rightarrow l}(X-\tau,X)+\frac{1}{2}\Psi^{(2)}_{l\rightarrow l}(X-\tau,X) \\
 = & \frac{1}{2\pi}e^{-i(k_{0}+p)(2X-\tau)}\left\{ \left[r_{r\rightarrow l}(p)+t_{l\rightarrow l}(-p)\right]^{2}e^{-ip\tau}+\left[2r_{r \rightarrow l}(p)+2t_{l\rightarrow l}(-p)-1-[r_{r\rightarrow l}(p)+t_{l\rightarrow l}(-p)]^{2}\right]\cos\left(\frac{\sqrt{\Delta}\tau}{2}\right)e^{-\Gamma\tau}\right.\nonumber \\
 & +\frac{\Gamma}{\Gamma-ip}\left[2-2r_{r\rightarrow l}(p)-t_{l\rightarrow l}(-p)-t_{r\rightarrow r}(p)\right]\left[\cos\left(\frac{\sqrt{\Delta}}{2}\tau\right)+\frac{(\alpha_{-}-\alpha_{+})\cos\theta-2i\beta\sin\theta}{\sqrt{\Delta}}\sin\left(\frac{\sqrt{\Delta}}{2}\tau\right)\right]e^{-\Gamma\tau}\nonumber \\
 +& \left.2\Gamma\frac{\left[\beta^{2}+i(\alpha_{-}\!-\!\alpha_{+})p\!-2\Gamma\alpha_{-}\right](1\!+\!\cos\theta)\!+2i(\Gamma -ip)\beta\sin\theta}{\sqrt{\Delta}D(p)}\sin\left(\frac{\sqrt{\Delta}\tau}{2}\right)[1-r_{r\rightarrow l}(p)\!-\!t_{l\rightarrow l}(-p)]e^{-\Gamma\tau}\right\},
\end{align}
is the superposition of the three two-photon functions for left-moving photons and
\begin{align}
\Psi_{r}^{(2)}&(X,X+\tau)  = \Psi^{(2)}_{{\rm TS}\rightarrow r}(X,X+\tau)+\frac{1}{2}\Psi^{(2)}_{l\rightarrow r}(X,X+\tau)+\frac{1}{2}\Psi^{(2)}_{r\rightarrow r}(X,X+\tau)\\
= & \frac{1}{2\pi}e^{i(k_{0}+p)(2X+\tau)}\left\{ \left[r_{l\rightarrow r}(-p)+t_{r\rightarrow r}(p)\right]^{2}e^{-ip\tau}+\left[2r_{l \rightarrow r}(-p)+2t_{r\rightarrow r}(p)-1-[r_{l \rightarrow r}(-p)+t_{r\rightarrow r}(p)]^{2}\right]\cos\left(\frac{\sqrt{\Delta}\tau}{2}\right)e^{-\Gamma\tau}\right.\nonumber \\
 & +\frac{\Gamma}{\Gamma-ip}\left[2-2r_{l \rightarrow r}(-p)-t_{l\rightarrow l}(-p)-t_{r\rightarrow r}(p)\right]\left[\cos\left(\frac{\sqrt{\Delta}}{2}\tau\right)-\frac{(\alpha_{-}-\alpha_{+})\cos\theta+2i\beta\sin\theta}{\sqrt{\Delta}}\sin\left(\frac{\sqrt{\Delta}}{2}\tau\right)\right]e^{-\Gamma\tau}\nonumber \\
 + & \left.2\Gamma\frac{\left[\beta^{2}\!+\!i(\alpha_{-}\!-\!\alpha_{+})p\!-2\Gamma\alpha_{-}\right](1\!+\!\cos\theta)-2i(\Gamma -ip)\beta\sin\theta}{\sqrt{\Delta}D(p)}\sin\left(\!\frac{\sqrt{\Delta}\tau}{2}\!\right)[1\!-\!r_{l \rightarrow r}(-p)\!-\!t_{l\rightarrow l}(-p)]e^{-\Gamma\tau}\right\},
\end{align}
is the superposition of the three two-photon functions for right-moving photons. Using the fact that $t_{l\rightarrow l}(\phi)=t_{r\rightarrow r}(-\phi)$, we can verify that
\begin{equation}
 g^{(2)}_l (\tau,\phi) = g^{(2)}_r (\tau,-\phi).   
\end{equation}

For the case $\tau = 0$, the $g^{(2)}$-functions reduce to
\begin{align}
g_{l}^{(2)}(0) & =\frac{\left|-ip\left[2r_{r \rightarrow l}(p)+2t_{l\rightarrow l}(-p)-1\right]+\Gamma[1+t_{l\rightarrow l}(-p)-t_{r\rightarrow r}(p)]\right|^{2}}{\left|r_{ r\rightarrow l}(p)+t_{l\rightarrow l}(-p)\right|^{4}|\Gamma-ip|^{2}},\\
g_{r}^{(2)}(0) & =\frac{\left|-ip\left[2r_{ l\rightarrow r}(-p)+2t_{r\rightarrow r}(p)-1\right]+\Gamma[1+t_{r\rightarrow r}(p)-t_{l\rightarrow l}(-p)]\right|^{2}}{\left|r_{l \rightarrow r}(-p)+t_{r\rightarrow r}(p)\right|^{4}|\Gamma-ip|^{2}}.   
\end{align}\end{widetext}
Specifically, for the resonant inputs with $p=0$, we have
\begin{equation}
g_{l}^{(2)}(0) =\frac{\left|D(0)\right|^{2}\left|D(0)+4ig\Gamma\sin\phi\sin\theta\right|^{2}}{\left|g^{2}+2g\Gamma e^{i\phi}\sin\theta+2i\Gamma(g\cos\phi+\Gamma\sin\theta)\right|^{4}},
\end{equation}
\begin{equation}
g_{r}^{(2)}(0)\!=\!\frac{\left|D(0)\right|^{2}\left|D(0)-4ig\Gamma\sin\phi\sin\theta\right|^{2}}{\left|g^{2}\!+\!2g\Gamma e^{-i\phi}\sin\theta\!+\!2i\Gamma(g\cos\phi\!+\!\Gamma\sin\theta)\right|^{4}},
\end{equation}
with
\begin{equation}
D(0)=g^{2}+(1-e^{2i\theta})\Gamma^{2}-2ig\Gamma e^{i\theta}\cos\phi.
\end{equation}
We can verify that $g_{\lambda}^{(2)}(0)\rightarrow 1$ when $g\rightarrow0$
for resonant drivings with $p=0$.

\begin{figure}
\includegraphics[width=8.5cm]{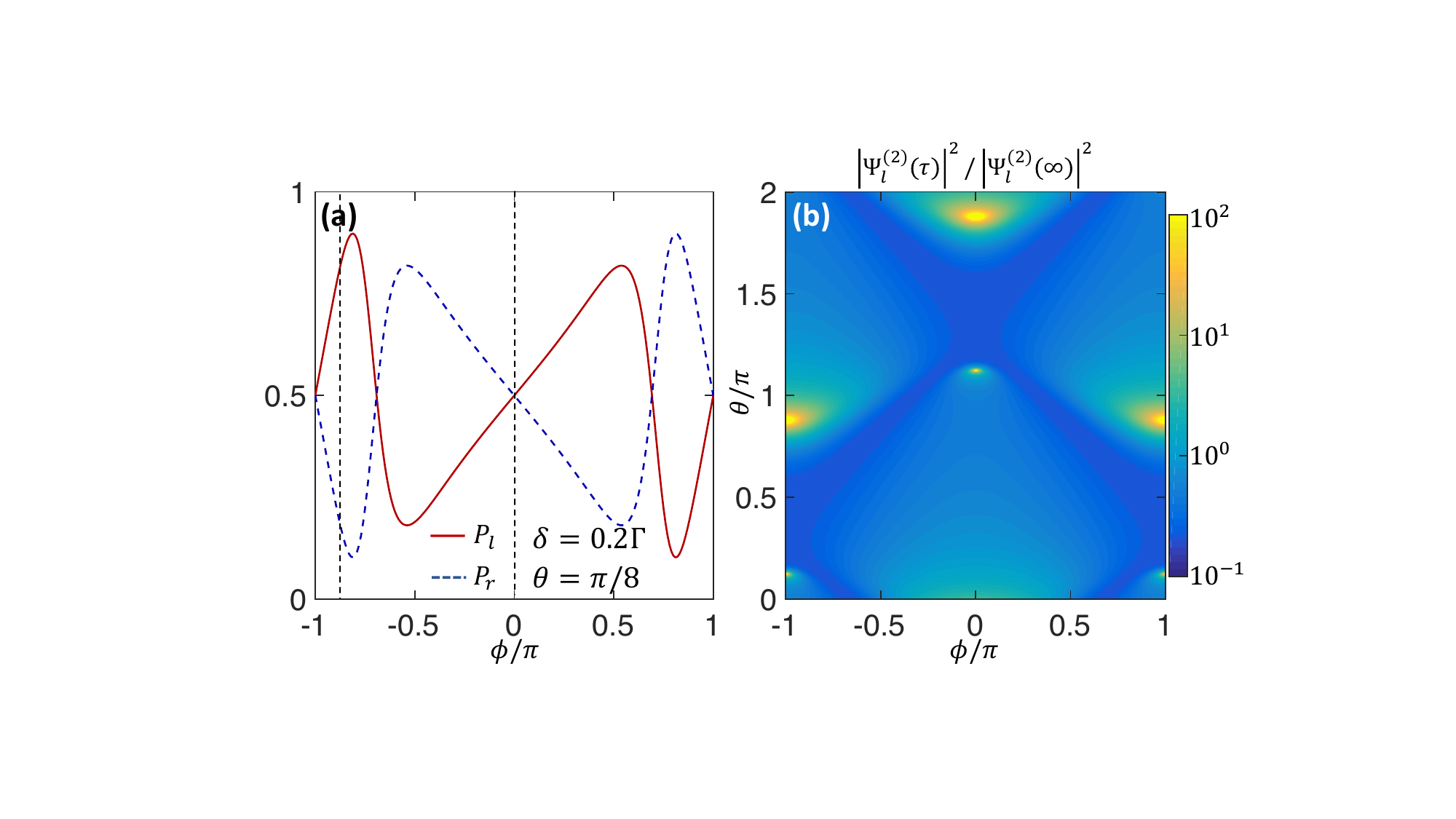}
\centering
\caption{\label{fig:S2} (a) The output probabilities of the left port (red solid curve) and the right port (blue dotted curve) for the two lines in Fig.~\ref{fig:3} (c) in the main text. (b) The effective $g^{(2)}(0)$-function for the Hong-Ou-Mandel interference case  $p=0.5\Gamma$.}
\end{figure}

The numerical simulation of the $g^{(2)}$-function is presented in the main text. Here, in Fig.~\ref{fig:S2} (a), we only display the output probabilities of the two ends for the two lines shown in Fig.~\ref{fig:3} (c). The output probabilities for the left and right ports at $\phi = -0.87\pi$ are around $P_l \approx 83\%$ and $P_r\approx 17\%$, respectively. For $\phi = 0$, the two output probabilities are the same, i.e., $P_l = P_r = 50\%$.

\section{Hong-Ou-Mandel interference\label{Appendix5}}
In this section, we investigate two-photon interference. Specifically, for the Hong-Ou-Mandel (HOM) interference, two incident photons will always exit from the same port for $50:50$ beam splitters. Traditional linear beam splitters do not alter the statistical properties of the input photons. However, this is not the case for our quantum beam splitter.

For a two-photon interference, there are three types of output states: (1) two photons both come out from the left output port; (2) two photons both come out from the right output port; (3) two photons come out from two different ports. The output probabilities are determined by output two-photon wave function $\Psi^{(2)}(X,X+\tau)$ with $X\gg \Gamma^{-1}$ and $\tau\rightarrow\infty$. The output two-photon functions for the first two cases have been obtained in Sec.~\ref{sec:TwoSides}. In evaluating the two-photon function $\Psi^{(2)}(X_1,X_2)$, we have assumed $|X_1|>|X_2|$. Our obtained two-photon wave functions are exchange symmetric, i.e. $\Psi^{(2)}_{{\rm TS}\rightarrow \lambda}(X_1,X_2) = \Psi^{(2)}_{{\rm TS}\rightarrow \lambda}(X_2,X_1)$. Thus, to obtain correct output probabilities, an extra factor $1/\sqrt{2}$ should be added due to the identity principle. The probabilities of the three output cases are determined by the following wave functions
\begin{align}
\left.\tilde{\Psi}_{{\rm TS}\rightarrow l}(X\!-\! \tau,X)\right|_{\tau\rightarrow\infty} & \!=\! \frac{e^{-i(k_{0}+p)(2X-\tau)}}{2\pi}\!\sqrt{2} r_{r \rightarrow l}(p)t_{l\rightarrow l}(-p), \\
\left.\tilde{\Psi}_{{\rm TS}\rightarrow r}(X\!+\!\tau,X)\right|_{\tau\rightarrow\infty} &\! =\! \frac{e^{i(k_{0}+p)(2X+\tau)}}{2\pi}\! \sqrt{2} r_{l \rightarrow r}(-p)t_{r\rightarrow r}(p), 
\end{align}
and the wave function for the third case
\begin{align}
& \left.\tilde{\Psi}_{\rm TS\rightarrow TS}(X+\tau,X)\right|_{\tau\rightarrow\infty}\nonumber\\
= & \frac{1}{2\pi}e^{i(k_{0}+p)\tau} \left[r_{r \rightarrow l}(p)r_{l \rightarrow r}(-p)+t_{l\rightarrow l}(-p)t_{r\rightarrow r}(p) \right].  
\end{align}
Using the relations (\ref{eq:rt_relation1}) and (\ref{eq:rt_relation2}) between  the scattering coefficients, we can verify the normalization condition,
\begin{align}
\left|\sqrt{2} r_{r \rightarrow l}(p) t_{l\rightarrow l}(-p)\right|^2 & + \left|\sqrt{2} r_{l \rightarrow r}(-p)t_{r\rightarrow r}(p)\right|^2\nonumber \\
+ |r_{r \rightarrow l}(p)& r_{l \rightarrow r}(-p)+t_{l\rightarrow l}(-p)t_{r\rightarrow r}(p)|^2 = 1. \label{eq:normalization}   
\end{align}
For HOM interference, both the reflection and transmission rates are $50\%$. In this case, we can verify that the third term in Eq.~(\ref{eq:normalization}) vanishes and two photons always come out from the same port.

The denominator of the $g^{(2)}$-function is characterized by the weighted average of the square of the photon numbers\begin{widetext}
\begin{align}
\overline{N^2_{l}} & = 2^2 \left|\sqrt{2} r_{r\rightarrow l}(p)t_{l\rightarrow l}(p)\right|^2 + \left|r_{r\rightarrow l}(p)r_{l\rightarrow r}(-p)+t_{l\rightarrow l}(-p)t_{r\rightarrow r}(p)\right|^2,\\
\overline{N^2_{r}} & = 2^2 \left|\sqrt{2} r_{l\rightarrow r}(-p)t_{r\rightarrow r}(p)\right|^2 + \left|r_{r\rightarrow l}(p)r_{l\rightarrow r}(-p)+t_{l\rightarrow l}(-p)t_{r\rightarrow r}(p)\right|^2.
\end{align}
\end{widetext}
The statistical properties of the two output photons can be characterized by an effective $g^{(2)}$-function
\begin{equation}
 g^{(2)}_{\lambda}(\tau) = \frac{\left|\Psi_{{\rm TS}\rightarrow \lambda}(X,X+\lambda \tau)\right|^2}{\overline{N^2_{\lambda}}/(2\pi)^2}. 
\end{equation}
For $\tau =0$, we have
\begin{equation}
g_{l}^{(2)}(0) =\frac{\left|1-\Gamma/(\Gamma-ip)\right|^{2}|r_{r\rightarrow l}(p)|^2}{\overline{N^2_{l}}},
\end{equation}
\begin{equation}
g_{r}^{(2)}(0) = \frac{\left|1-\Gamma/(\Gamma-ip)\right|^{2}|r_{l\rightarrow r}(-p)|^2}{\overline{N^2_{r}}} 
\end{equation}
Here, we observe that the function $g^{(2)}(0)$ is parity symmetric as shown in Fig.~\ref{fig:S2} (b), i.e., $g_{l}^{(2)}(0,\phi )=g_{r}^{(2)}(0,\phi) = g_{l}^{(2)}(0,-\phi)$. Furthermore, for resonant two-photon interference with $p=0$, the output photons are always sub-Poissonian with $g^{(2)} (0) = 0$. However, the $g^{(2)}(\tau)$ functions of the photons from the left and right ports behave differently. Consequently, the parity symmetry of the $g^{(2)}(\tau)$ function has been broken.

\section{Experimental implementation\label{Appendix6}}
In this section, we provide detailed insights into implementing our proposed theoretical model within circuit QED systems. Without loss of generality, We consider two planar transmon qubits coupling to a coplanar waveguide~\cite{Joshi2023Resonance,Kannan2023Ondemand}. The Hamiltonian governing the entire system is represented by $\hat{H}^{\prime}=\hat{H}_a^{\prime}+\hat{H}_p+\hat{H}_{\rm int}^{\prime}$. Two artificial atoms under modulation are described by 
\begin{align}
\hat{H}_{a}^{\prime}= & \omega_{0}\hat{\sigma}_{1}^{\dagger}\hat{\sigma}_{1}+\left(\omega_{0}+\Delta-A\cos\Delta t\right)\hat{\sigma}_{2}^{\dagger}\hat{\sigma}_{2}\nonumber\\
&+g^{\prime}\cos\left(\Delta t+\phi^{\prime}\right)\left(\hat{\sigma}_{1}^{\dagger}\hat{\sigma}_{2}+{\rm H.c.}\right),
\end{align}
where $\Delta$ is the frequency difference between two qubits. To achieve the parity-symmetry breaking atom-atom interaction (\ref{eq:atom_interaction}), we employ two phase-locked microwave drives operating at the identical frequency $\Delta$ and tunable phase difference $\phi^{\prime}$ to modulate the frequency of the qubit-2 and the qubit-qubit coupling, respectively. The frequency and coupling modulation methods have been routinely used in circuit systems~\cite{roushan2017chiral,McKay2016Universal,Wu2018,Roth2017Analysis}. The waveguide photon Hamiltonian is the same as $\hat{H}_{p}=\int dk(\omega_{0}+\lambda k)\hat{b}_{k,\lambda}^{\dagger}\hat{b}_{k,\lambda}$. The coupling of the qubits to the waveguide is described by
\begin{equation}
\hat{H}_{{\rm int}}^{\prime}=\sum_{i}\eta_{i}\hat{\sigma}_{i}\int dk\left[\hat{b}_{k,r}^{\dagger}e^{-i(k_{0}+k)x_{i}}+\hat{b}_{k,l}^{\dagger}e^{i(k_{0}-k)x_{i}}\right],
\end{equation}
where the strengths are denoted by $\eta_1$ and $\eta_2$ respectively.

Under the condition $|\Delta|\gg A, g^\prime$, we can simplify the system Hamiltonian. In the rotating frame with respect to $\omega_{0}\hat{\sigma}_{1}^{\dagger}\hat{\sigma}_{1}+\left(\omega_{0}+\Delta-A\cos\Delta t\right)\hat{\sigma}_{2}^{\dagger}\hat{\sigma}_{2}+\omega_{0}\int dk\hat{b}_{k,\lambda}^{\dagger}\hat{b}_{k,\lambda}$,
we have~\cite{Ashhab2007Twolevel,zhou2008quantum,clerk2022introduction} 
\begin{align}
\hat{H}_{a}^{\prime} & =g^{\prime}\cos\left(\Delta t+\phi^{\prime}\right)\left[\hat{\sigma}_{1}^{\dagger}\hat{\sigma}_{2}\exp\left(-i\Delta t+i\frac{A}{\Delta}\sin\Delta t\right)+{\rm H.c.}\right]\nonumber\\
 & \approx\frac{1}{2}g^{\prime}\left[J_{0}\left(\frac{A}{\Delta}\right)e^{i\phi^{\prime}}+J_{2}\left(\frac{A}{\Delta}\right)e^{-i\phi^{\prime}}\right]\hat{\sigma}_{1}^{\dagger}\hat{\sigma}_{2}+{\rm H.c.} \\
& \equiv ge^{i\phi}\hat{\sigma}_{1}^{\dagger}\hat{\sigma}_{2}+{\rm H.c.},
\end{align}
where $J_n(x)$ is the $n$-th Bessel function of the first kind, $g$ and $\phi$ are the modulus and argument of $g^{\prime}\left[J_{0}\left(\frac{A}{\Delta}\right)e^{i\phi^{\prime}}+J_{2}\left(\frac{A}{\Delta}\right)e^{-i\phi^{\prime}}\right]/2$,
respectively, and we have used the Jacobi–Anger expansion and neglected non-resonant terms in the second step. Here, we have successfully achieved the non-symmetric atom-atom coupling we aimed for. Both its strength $g$ and phase $\phi$ can be finely tuned in experiments. 

The coupling between atom-1 and the waveguide photon remains unchanged. However, the coupling between atom-2 and the photons undergoes changes
\begin{align}
\hat{H}_{{\rm int,2}}^{\prime}  = &\eta_{2}\exp\left(-i\Delta t+i\frac{A}{\Delta}\sin\Delta t\right)\nonumber\\
& \times\int dk\hat{\sigma}_{2}\left[\hat{b}_{k,r}^{\dagger}e^{-i(k_{0}+k)x_{2}}+\hat{b}_{k,l}^{\dagger}e^{i(k_{0}-k)x_{2}}\right]\\
 \approx &  \eta_{2}J_{1}\left(\frac{A}{\Delta}\right)\int dk\hat{\sigma}_{2}\left[\hat{b}_{k,r}^{\dagger}e^{-i(k_{0}+k)x_{2}}+\hat{b}_{k,l}^{\dagger}e^{i(k_{0}-k)x_{2}}\right].
\end{align}
By tuning $A$ and $\Delta$, such that $\eta_{2}J_{1}(A/\Delta)=\eta_{1}=\eta$,
the interaction Hamiltonian $\hat{H}_{{\rm int}}^{\prime}$ reduces
to $\hat{H}_{{\rm int}}$ in Eq.~(\ref{eq:atom_interaction}).

\bibliography{main}
\end{document}